\documentclass[prd, aps, nofootinbib,preprintnumbers, showpacs,superscriptaddress,twocolumn]{revtex4}
\usepackage{latexsym}
\usepackage{amssymb}
\usepackage{amsfonts}
\usepackage{amsmath}
\usepackage{bm}
\usepackage[dvips]{graphicx}
\usepackage{color}

\newcommand{\be}{\begin{equation}}  
\newcommand{\ee}{\end{equation}}
\newcommand{\bea}{\begin{eqnarray}}           
\newcommand{\eea}{\end{eqnarray}} 
\newcommand{\beqn}{\begin{eqnarray*}}
\newcommand{\eeqn}{\end{eqnarray*}}
\newcommand{\ba}{\begin{align}}
\newcommand{\ea}{\end{align}}

\def\de{\partial}

\def\l{{\ell }}

\def\E{{\cal E}}
\def\B{{\cal B}}

\def\k{\kappa}
\def\a{\alpha}
\def\b{\beta}

\begin{document}

%%%%%%%%%%%%%%%%%%%%%%%%%%%%%%%%%%%%%%%%%%%%%%%%%%%%%%%%%%%%%%%%%%%%%%%

\title{Effective One Body description of tidal effects in inspiralling 
compact binaries}

%%%%%%%%%%%%%%%%%%%%%%%%%%%%%%%%%%%%%%%%%%%%%%%%%%%%%%%%%%%%%%%%%%%%%%%
\author{Thibault \surname{Damour}}
\affiliation{Institut des Hautes Etudes Scientifiques, 91440 Bures-sur-Yvette, France}
\affiliation{ICRANet, 65122 Pescara, Italy}

\author{Alessandro \surname{Nagar}}
\affiliation{Institut des Hautes Etudes Scientifiques, 91440 Bures-sur-Yvette, France}
\affiliation{ICRANet, 65122 Pescara, Italy}

\begin{abstract}
The late part of the gravitational wave signal of binary neutron star inspirals 
can in principle yield crucial information on the nuclear equation of state via its
dependence on relativistic tidal parameters. In the hope of analytically
describing the gravitational wave phasing during the late inspiral
(essentially up to contact)  we propose
an extension of the effective one body (EOB) formalism which includes tidal effects.
We compare the prediction of this tidal-EOB formalism to recently computed 
nonconformally flat quasi-equilibrium circular sequences of binary neutron 
star systems. Our analysis suggests the importance of higher-order (post-Newtonian)  
corrections to tidal effects, even beyond the first post-Newtonian order, and their
tendency to {\it significantly} increase the ``effective tidal
polarizability'' of neutron stars. 
We compare the EOB predictions to some recently advocated, nonresummed,
post-Newtonian based (``Taylor-T4'') description of the phasing of inspiralling systems.
This comparison shows the strong sensitivity of the late-inspiral phasing
to the choice of the analytical model, but raises the hope that a sufficiently
accurate numerical--relativity--``calibrated'' EOB model might give us a reliable 
handle on the nuclear equation of state.

\end{abstract}

\date{\today}


\maketitle

\section{Introduction}
\label{sec:sec1}

Some of the prime targets of the currently operating network of ground-based detectors of 
gravitational  waves (GWs) are the signals emitted by inspiralling and coalescing compact binaries. 
Here, ``compact binary'' refers to a binary system made either of two black holes, a black hole and a 
neutron star, or two neutron stars. The GW signal emitted by binary black hole (BBH) systems has 
been the subject of intense theoretical studies, based either on analytical methods or on numerical 
ones. In particular, recent progress in the application of the effective one body (EOB) approach to 
BBH systems has led to a remarkable agreement between the (analytical) EOB predictions and the 
best current numerical relativity results~\cite{Damour:2009kr,Buonanno:2009qa}
(see also~\cite{Yunes:2009ef}).
By contrast, much less work has been devoted to the study of 
the GW signal emitted by compact binaries comprising neutron stars: either black-hole-neutron-star 
(BHNS) systems or binary neutron-star (BNS) ones. During the inspiral phase (before contact), these 
systems differ from the BBH ones by the presence of tidal interactions which affect both the dynamics 
of the inspiral and the emitted waveform. During the merger and coalescence phase, the presence of 
neutron stars drastically modifies the GW signal~\cite{Baiotti:2009gk,Giacomazzo:2009mp,Baiotti:2008ra}. 
The coalescence signal involves (especially in the 
BNS case) a lot of  complicated physics and astrophysics, and is, probably, not amenable to the type 
of accurate analytical  description which worked in the BBH case. Early works
on this problem have tried to approximately  relate some qualitative features
of the merger GW
signal linked, e.g., to ``tidal disruption'', to  analytically describable
inputs~\cite{Bildsten:1992my,Kochanek:1992wk,Vallisneri:1999nq}.
 
Recently, Flanagan and Hinderer~\cite{Flanagan:2007ix,Hinderer:2007mb,Hinderer:2009} have
initiated the program of studying the quantitative  influence of
tidal effects~\cite{Hinderer:2007mb,Damour:2009vw,Binnington:2009bb} 
in inspiralling BNS systems. 
However, they only considered the early (lower frequency) portion of 
the GW inspiral signal, mainly because they were using a post-Newtonian
based description of the binary dynamics whose validity is restricted to
low enough frequencies.
In particular, one of the results of the recent work of 
Hinderer et al.~\cite{Hinderer:2009} is to show that the accumulated
GW phase due to tidal interactions is, for most realistic NS models of
mass $M\sim 1.4M_\odot$ {\it smaller} than the ``uncertainty''  in
the PN-based description of GW phasing (see the central panel of
their Fig.~4 where the thin-dashed and thin-dotted lines are two
measures of the PN ``uncertainty''. [These measures are larger than
the inspiral tidal signal except for the extreme case where the 
radius of the $1.4M_\odot$ NS is taken to be $\geq 16$ km].

By contrast, our aim in this work will be to propose a way of describing
the binary dynamics (including tidal effects) whose validity does not
have the limitations of PN-based descriptions and therefore
is not apriori limited
to the low frequency part, but extends to significantly higher 
frequencies. This might be crucial to increase the detectability of the
GW signal and thereby have a handle on the nuclear equation of state (EOS).
Indeed, our proposal consists in extending the EOB method
by incorporating tidal effects in it. 
Our hope is that such a tidally-extended EOB framework will be able to describe with 
sufficient approximation not only the early inspiral phase, but also the late inspiral up to the moment 
(that we shall consistently determine within our scheme) of ``contact''. 
We think that the present EOB description of tidal effects is likely to be
 more accurate than any of the possible 
``post-Newtonian-based'' descriptions involving supplementary tidal terms 
(such as~\cite{Flanagan:2007ix} or~\cite{Hinderer:2009}). 
This should be especially true in the BHNS systems which, in the limiting case 
$m_{\rm NS} \ll m_{\rm BH}$, are known to be well described by the EOB
approach (and rather badly described by post-Newtonian-based approaches).
We will give some evidence of the validity of the EOB description of close
neutron star systems by comparing our analytical predictions to recently
calculated quasi-equilibrium neutron star (NS) sequences of circular
orbits~\cite{Uryu:2009ye} (see also~\cite{Uryu:2005vv}).

%----------------------------------------------------------------------------------------------
\section{Effective-action description of tidal effects in two-body systems}
\label{sec:sec2}
%----------------------------------------------------------------------------------------------

\subsection{General formalism}
\label{sec:general}

The general relativistic tidal properties of neutron stars have been recently
studied in Refs.~\cite{Hinderer:2007mb,Damour:2009vw,Binnington:2009bb,Hinderer:2009}. 
 As emphasized in~\cite{Damour:2009vw}, there are (at least) three different 
types of tidal responses of a neutron star 
to an external tidal solicitation, which are measured by three different tidal
coefficients: (i) a gravito-electric-type coefficient 
$G\mu_\ell = [{\rm length}]^{2\ell +1}$ measuring the $\l^{\rm th}$-order 
mass multipolar moment $G M_{a_1\dots a_\l}$ induced in a star by an external
$\l^{\rm th}$-order gravito-electric tidal field $G_{a_1,\dots,a_\l}$; (ii)
a gravito-magnetic-type coefficient $G\sigma_\l=[{\rm length}]^{2\l+1}$ 
measuring the $\l^{\rm th}$ spin multipole moment $G S_{a_1 \dots a_\l}$
induced in a star by an external $\l^{\rm th}$-order gravito-magnetic tidal
field $H_{a_1\dots a_\l}$; and (iii) a dimensionless ``shape'' Love number
$h_\ell$ measuring the distorsion of the shape of the surface of a star by an
external $\l^{\rm th}$-order gravito-electric tidal field. It was found in
\cite{Damour:2009vw,Binnington:2009bb} that all those coefficients have a strong sensitivity to the
value of the star's ``compactness'' $c\equiv GM/c_0^2 R$ (where we denote by
$c_0$ the velocity of light, to be distinguished from the compactness $c$).
This means, in particular, that the numerical values of the tidal coefficients
of NS's should not be evaluated by using Newtonian estimates. Indeed, the
dimensionless version of $\mu_\ell$, traditionally denoted as $k_\ell$
(``second Love number'') and defined as
\be
\label{eq:kl}
2 k_\l \equiv (2\l -1)!! \dfrac{G\mu_\l}{R^{2\l +1}} ,
\ee
where $R$ denotes the areal radius of the NS, is typically three times smaller
than its Newtonian counterpart (computed from the same equation of state). A
similar, though less drastic, ``quenching'' also occurs for the ``first Love
number'' $h_\l$. In particular, though Newtonian $h_\l$'s are larger than $1$
(and equal to $1+2k_\l$, see Eq.~(81) of \cite{Damour:2009vw}), the typical relativistic
values of $h_\l$ are smaller than $1$. This will play a useful role in our
analysis below of the moment where the tidal distortion of the NS becomes too
large for continuing to use an analytical approach. 

It was shown in \cite{Damour_cras80, Damour1983} that the 
motion and radiation of two black holes can be described, up 
to the fifth post-Newtonian (5PN) approximation, by an 
effective action of the form
\be
\label{eq:2.1n}
S_0 = \int d^D x \, \dfrac{\sqrt{g} \, R(g)}{16\pi \, G} + S_{\rm point\mbox{-}mass},
\ee
where
\be
\label{eq:2.2}
S_{\rm point\mbox{-}mass} = -\sum_A \int M_A \, ds_A ,
\ee
is a ``skeletonized'' description of black holes, as ``point masses''. 
To give meaning to the addition of point-mass sources to the nonlinear 
Einstein equations, one needs to use a covariant regularization
method. Refs.~\cite{Damour_cras80, Damour1983} mainly used 
Riesz' analytic regularization, but it was already mentioned at the time that 
one could equivalently use dimensional regularization. The efficiency 
and consistency of the latter method was shown by the calculations of the dynamics, 
and radiation, of BBH systems at the 3PN
level~\cite{Damour:2001bu,Blanchet:2003gy,Blanchet:2004ek}. Let us also 
recall that the limitation to the 5PN 
level in Ref.~\cite{Damour1983} is precisely linked to the possible appearance of ambiguities in BBH 
dynamics appearing at the level where tidal effects start entering the picture. 
Indeed, it is well-known in effective field theory that finite-size effects correspond
to augmenting the point-mass action~\ref{eq:2.1n} by non-minimal (worldline) 
couplings involving higher-order derivatives of the field 
[see~\cite{Damour:1995kt,Goldberger:2004jt} and Appendix~A of Ref.~\cite{Damour:1998jk}].
More precisely, 
the two tidal effects parametrized by $\mu_\l$ and $\sigma_\l$ correspond to augmenting the leading 
point-particle effective action, (\ref{eq:2.1n}), (\ref{eq:2.2}), by the following nonminimal worldline
couplings
\begin{eqnarray}
\label{eq:2.3}
\Delta S_{\rm nonminimal} &= &\sum_A \biggl\{ \frac{1}{2} \, \frac{1}{\l!} \, \mu_\l^A \int ds_A (G_L^A)^2 
\nonumber \\
&+ &\frac{1}{2} \, \frac{\l}{\l + 1} \, \frac{1}{\l!} \, \frac{1}{c_0^2} \, \sigma_\l^A \int ds_A (H_L^A)^2 \biggl\} .
\end{eqnarray}
Here\footnote{We use here the notation of~\cite{Damour:1990pi}, notably for multi-indices $L \equiv a_1 , \ldots , a_\l$.} 
$G_L^A \equiv G_{a_1 \ldots a_\l}^A$ and $H_L^A \equiv H_{a_1 \ldots a_\l}^A$ are the 
gravito-electric and gravito-magnetic ``external'' tidal gradients evaluated along the worldline of the 
considered star (labelled by $A$), in the local frames (attached to body $A$) defined in~\cite{Damour:1990pi}. 
If needed, they can be reexpressed in terms of covariant derivatives of the Riemann (or Weyl) 
tensor. For instance, using Eq.~(3.40) of~\cite{Damour:1990pi}, the leading, 
quadrupolar terms in Eq.~(\ref{eq:2.3}) read
\begin{eqnarray}
\label{eq:2.4} 
\Delta S_{\rm nonminimal} &= &\sum_A \biggl\{ \frac{1}{4} \, \mu_2^A \int ds_A \, {\mathcal E}_{\a\b}^A 
\, {\mathcal E}^{A_{\a\b}} \nonumber \\
&+ &\frac{1}{6} \, \sigma_2^A \int ds_A \, {\mathcal B}_{\a\b}^A \, 
{\mathcal B}^{A_{\a\b}} + \cdots \biggl\}
\end{eqnarray}
where $\E_{\a\b}^A \equiv [u^{\mu} \, u^{\nu} \, C_{\mu\a\nu\b}]^A$, $\B_{\a\b}^A \equiv [u^{\mu} \, 
u^{\nu} \, C_{\mu\a\nu\b}^*]^A$, with $C_{\mu\nu\a\b}^* \equiv \frac{1}{2} \, \epsilon_{\mu\nu\rho
\sigma} \, C^{\rho\sigma}_{\a\b}$ being the dual of the Weyl tensor $C$, and $u^{\mu} = dz^{\mu} / ds$ 
being the four-velocity along the considered worldline. As explained in Appendix A of 
Ref.~\cite{Damour:1998jk}, one can, modulo some suitable ``field redefinitions'' that do 
not affect the leading result, indifferently use the Weyl tensor 
$C_{\a\b\mu\nu}$ or the Riemann tensor $R_{\a\b\mu\nu}$ in evaluating the $\E_{\a\b}$ and 
$\B_{\a\b}$ entering Eq.~(\ref{eq:2.4}).

The effective-action terms (\ref{eq:2.3}), (\ref{eq:2.4}) can be used to compute the various observable 
 effects linked to the relativistic tidal coefficients $\mu_\l$ and $\sigma_\l$~\footnote{More precisely, 
 Eq.~(\ref{eq:2.3}) describes only the effects that are {\it linear} in tidal
 deformations (and which preserve {\it parity}). If one wished to also 
 consider {\it nonlinear} tidal effects one should augment the {\it quadratic-only} terms (\ref{eq:2.4}) by 
 higher-order nonminimal worldline couplings which are cubic, quartic, etc$\ldots$ in $C_{\mu\a\nu\b}$ and 
 its gradients. The coefficients of such terms would then parametrize some {\it nonlinear} tidal effects, 
 which have not been considered in the linear treatments of Refs.~\cite{Damour:2009vw,Binnington:2009bb}.}. 
 In particular,  they imply both: (i) additional terms in the dynamics of the considered binary system, and 
 (ii) additional terms in the gravitational radiation emitted by the considered binary system. 
 Both types of additional terms can, in principle, be evaluated with any needed relativistic accuracy 
 from Eq.~(\ref{eq:2.3}), i.e. computed either in a ``post-Minkowskian'' (PM) expansion in powers of 
 $G/c_0^2$, or (after a further re-expansion in powers of $1/c_0$), in a ``post-Newtonian'' (PN) 
 expansion in powers of $1/c_0^2$. Let us remark in passing that the PM expansion can be 
 conveniently expressed in terms of Feynman-like diagrams, as was explicitly discussed (for 
 tensor-scalar gravity) at the 2PN level in~\cite{Damour:1995kt}.

 Here we shall use the extra terms 
 (\ref{eq:2.3}), (\ref{eq:2.4}) as a way to {\it add} to the description of binary black hole systems 
 the effects linked to the replacement of one or two of the black holes by a neutron star. From this 
 point of view, we shall conventionally consider that the tidal coefficients of a black hole vanish: 
 $\mu_\l^{\rm BH} = 0 = \sigma_\l^{\rm BH}$~\cite{Damour:2009vw,Binnington:2009bb}.
 However, as emphasized in~\cite{Damour:2009vw}, 
 more work is needed to clarify whether this is exact, i.e. whether the description of BBH's 
 by an effective action requires or not the presence of additional couplings of the 
 type of Eqs.~(\ref{eq:2.3}), (\ref{eq:2.4}), as ``counter terms'' to absorb
dimensional regularization poles  $\propto (D-4)^{-1}$ (such poles are indeed linked
to the possible ambiguities expected to arise at 5PN in the point-mass dynamics; 
see the discussion in Sec.~5
 of~\cite{Damour1983}; see also Sec.~7 of~\cite{Damour:2009sm}) . 
 We leave to future work a clarification of this subtle issue.

\subsection{Leading-Order tidal effects in the two-body interaction Lagrangian}
 
Let us first consider the {\it dynamical} effects, implied by (\ref{eq:2.3})
i.e. the  tidal contribution to the ``Fokker'' Lagrangian describing 
the dynamics of two compact bodies after having integrated out the gravitational field, say 
\be
\label{eq:2.5}
L ({\bm q}^A , {\bm v}^A) = L^{\rm point\mbox{-}mass} + L^{\rm tidal} \, .
\ee
Here, $L^{\rm point\mbox{-}mass} (q,v)$ denotes the (time-symmetric) interaction Lagrangian 
following from the point-mass action (\ref{eq:2.1n}) (say after a suitable redefinition of position variables 
to eliminate higher derivatives). It is currently known at the 3PN level. The supplementary term 
$L^{\rm tidal}$ in Eq.~(\ref{eq:2.5}) is of the symbolic form (keeping only powers of $G$ and $1/c_0$) 
\begin{eqnarray}
\label{eq:2.5bis}
L^{\rm tidal} &\sim &G^2 \, \mu_2 \left( 1 + \frac{1}{c_0^2} + G + \cdots \right)  \nonumber \\
&+ &\frac{G^2 \, \sigma_2}{c_0^2} \, \left( 1 + \frac{1}{c_0^2} + G + \cdots \right) \nonumber \\
&+ &G^2 \, \mu_3 \left( 1 + \frac{1}{c_0^2} + G + \cdots \right) + \cdots
\end{eqnarray}
Let us start by discussing the {\it leading order} contributions associated to each
tidal coefficient $\mu_\ell$ or $\sigma_\ell$.
The leading term in the contribution linked to $\mu_\l$ is simply obtained from (\ref{eq:2.3}) by 
inserting the leading-order value of $G_L^A$, i.e. $(L \equiv a_1 \ldots a_\l)$
\be
\label{eq:2.6}
G_L^A = \left[ \partial_L U^{\rm ext }( {\bf x} ) \right]^A = \partial_L^A 
\left( \frac{GM^B}{\vert {\bm z}_A - {\bm z}_B \vert} \right)
\ee
where $B \ne A$ denotes the companion of body $A$ in the considered binary system ($A,B=1,2$), 
and $\vert {\bm z}_A - {\bm z}_B \vert$ the distance between the two bodies. In addition 
$\partial_L^A \equiv \partial_{a_1 \ldots a_\l}^A$, with $\partial_a^A \equiv \partial / \partial z_A^a$, 
denotes the differentiation with respect to ${\bm z}_A$ that appear after taking the limit where 
the field point ${\bm x}$ tends to ${\bm z}_A$ on the worldline of body $A$. Using
\be
\label{eq:2.7}
\partial_L^A \, \frac{1}{r_{AB}} = (-)^\l \, (2\l - 1)!! \, \frac{\hat n_{AB}^L}{r_{AB}^{\l + 1}}
\ee
where $n_{AB}^a \equiv (z_A^a - z_B^a) / r_{AB}$, $r_{AB} \equiv \vert {\bm z}_A - {\bm z}_B \vert$, 
and where the hat denotes a symmetric trace-free (STF) projection, and the fact that 
(see, e.g., Eq.~(A25) of~\cite{BD86})
\be
\label{eq:2.8}
\hat n^L_{AB}\, \hat n^L_{AB} = \hat n^L_{AB} \, n^L_{AB} = \frac{\l !}{(2\l - 1)!!} \, ,
\ee
one easily finds that the leading Lagrangian contribution proportional to $\mu_\l$ reads
\begin{eqnarray}
\label{eq:2.9}
L_{\mu_\l^A} &= &\frac{(2\l-1)!!}{2} \, \mu_\l^A \, \frac{(GM^B)^2}{r_{AB}^{2\l+2}} \nonumber \\
&= &k_\l^A \, G(M^B)^2 \, \frac{R_A^{2\l + 1}}{r_{AB}^{2\l+2}} \, .
\end{eqnarray}
Here we have used (\ref{eq:kl}) to replace $G\mu_\l^A$ in terms of the dimensionless Love number 
$k_\l^A$, and of the areal radius $R_A$ of the NS. Note that, in a BNS system, one has to add two 
different contributions: $L_{\mu_\l^A} + L_{\mu_\l^B}$. By contrast, in a BHNS system one has only 
$L_{\mu_\l^A}$ if $A$ denotes the NS.

Let us also evaluate the leading ``magnetic-type'' contribution, i.e. the term $\propto \sigma_2$ in 
(\ref{eq:2.5}). It is obtained by inserting in (\ref{eq:2.3}) the ``Newtonian''-level value of the 
gravito-magnetic quadrupolar field $H_{ab}^{B/A}$ exterted by body $B$ on body $A$. This is given 
by Eq.~(6.27a) of \cite{Damour:1991yw}, namely
\begin{align}
\label{eq:2.10}
H_{ab}^{B/A} =& -2G \, \partial_{ac}^A \left( \frac{\epsilon_{bcd} \, M^B \, v_{BA}^d}{r_{AB}} \right) \nonumber \\
&-2G \, \partial_{bc}^A \left( \frac{\epsilon_{acd} \, M^B \, v_{BA}^d}{r_{AB}} \right) 
\end{align}
where $v_{BA}^d \equiv v_B^d - v_A^d$ is the relative velocity between $B$ and $A$. 
A straightforward calculation then yields
\be
\label{eq:2.11}
L_{\sigma_2^A} = 12 \, \sigma_2^A \, \frac{(GM^B)^2}{r_{AB}^6} \left[ \left( \frac{{\bm v}_{AB}}{c_0} 
\right)^2 - \left( \frac{{\bm n}_{AB} \cdot {\bm v}_{AB}}{c_0} \right)^2 \right] \, .
\ee

Note that the leading quadrupolar gravito-magnetic contribution (\ref{eq:2.11}) is 
smaller than the corresponding quadrupolar gravito-electric contribution
\be
\label{eq:2.12}
L_{\mu_2^A} = \frac{3}{2} \, \mu_2^A \, \frac{(GM^B)^2}{r_{AB}^6}
\ee
by a factor
\be
\label{eq:2.13}
8 \, \frac{\sigma_2^A}{\mu_2^A} \left[ \left( \frac{{\bm v}_{AB}}{c_0} \right)^2 - \left( \frac{{\bm n}_{AB} 
\cdot {\bm v}_{AB}}{c_0} \right)^2 \right] \, .
\ee
In terms of the corresponding dimensionless Love numbers $j_2$ (defined in \cite{Damour:2009vw}) 
and $k_2$,  the prefactor $8 \, \sigma_2^A / \mu_2^A$ is equal to the dimensionless ratio $j_2 / (4k_2)$. 
However, it was found in \cite{Damour:2009vw,Binnington:2009bb} that the magnetic Love number $j_2$ was much smaller 
than $k_2$. Typically, for a $\gamma = 2$ $\mu$-polytrope and a compactness $c^A \sim 0.15$, 
one has $j_2 \simeq - 0.02$, while $k_2 \sim 0.1$, so that $8 \, \sigma_2 / \mu_2 = j_2 / (4k_2) 
\simeq -0.05$. In other words, the leading gravito-magnetic interaction (\ref{eq:2.11}) is equivalent 
(say for circular orbits) to a 1PN fractional correction factor, $1+\a \, (v_{AB} / c_0)^2$, modifying 
the leading gravito-electric contribution (\ref{eq:2.12}), with $\a = 8 \, \sigma_2 / \mu_2 = j_2 / (4k_2) 
\sim -0.05$. As we shall discuss below,  the 1PN correction to (\ref{eq:2.12}), 
implied by \eqref{eq:2.3}, involves coefficients $\a^{\rm 1PN}$ of order unity.
We will therefore, in the following, neglect the 
contribution (\ref{eq:2.11}) which represents only a small fractional
modification to the 1PN correction to (\ref{eq:2.12}).
On the other hand, we shall retain some of the higher-degree gravito-electric contributions.
Indeed, though, for instance, $L_{\mu_3^A}\propto 1/r_{AB}^8$ formally
corresponds to a 2PN correction to $L_{\mu_2^A}\propto 1/r_{AB}^6$, its
coefficient is much larger than that corresponding to an order-unity 2PN
correction to Eq.~\eqref{eq:2.12} [see Table~\ref{tab:table1} below].

Summarizing: the leading-order tidal contributions to the two-body interaction Lagrangian are 
(from Eq.~(\ref{eq:2.9}))
\be
\label{eq:2.14}
L^{\rm tidal} = + G \sum_{\l \geq 2} \left\{ k_\l^A (M^B)^2 \, \frac{R_A^{2\l+1}}{r_{AB}^{2\l+2}} + k_\l^B 
(M^A)^2 \, \frac{R_B^{2\l+1}}{r_{AB}^{2\l+2}} \right\} \, ,
\ee
where $k_\l^A$ denotes the $\l^{\rm th}$ dimensionless Love number of a NS \cite{Hinderer:2007mb,Damour:2009vw,Binnington:2009bb}. 
Note that the plus sign in Eq.~(\ref{eq:2.14}) expresses the fact that the tidal interactions are {\it attractive}.

\subsection{Structure of subleading (post-Newtonian) dynamical tidal effects}

Leaving to future work~\cite{DEF09} a detailed computation of higher-order 
relativistic tidal effects, let us indicate their general structure. Here, we
shall neglect the effects which are {\it nonlinear} in the worldline
couplings $\mu^A_\ell$ of Eq.~\eqref{eq:2.3} (e.g. effects $\propto \mu_2^A \mu_2^A$)
for two reasons. On the one hand, such effects are numerically quite small,
even for close neutron stars (as we shall check below). On the other hand, a 
fully consistent discussion of such effects requires that one considers
a more general version of nonminimal worldline couplings, involving terms which
are cubic (or more nonlinear) in the curvature tensor and its covariant derivatives.
Indeed, it is easily seen that a nonminimal coupling which is {\it cubic} in
$G_{ab}\sim {\cal E}_{\alpha\beta}$ contributes to the dynamics at the same
level that a 1PN correction to the coupling quadratic in $G_{abc}$.

In the "quadratic-in-curvature" approximation of Eq.~\eqref{eq:2.3} the part of the tidal
interaction which is proportional to $\mu_\ell^A$ will have the symbolic structure
\begin{align}
\label{symb}
S_{\mu^A}\sim \mu^A(G M^B)^2\bigg[1&+GM^A + GM^B \nonumber\\
                             &+ \left(GM^A+GM^B\right)^2+\dots \bigg]
\end{align}
where we only indicate the dependence on $GM^A$ and $GM^B$, leaving 
out all the coefficients (symbolically replaced by 1), which depend on positions and
velocities. The presence of an overall factor $(GM^B)^2$ comes from the fact that
$G^A_\ell (z^\mu)$ in Eq.~\eqref{eq:2.3} (which denotes the {\it regularized} value of 
some gradient of the curvature tensor as the field point $x$ tends to $z_A^\mu(s_A)$ 
on the worldline of $M^A$) is proportional to $GM^B$, so that it is vanishing 
when $M^B\to 0$, i.e. in the limit of a one-body system.
[We are considering here a two-body system; in the more general case of an
  $N$-body system we would have $G^A(z_A)\propto \sum_{B\neq A} G M^B$.] 
In a diagrammatic language (see e.g.~\cite{Damour:1995kt}) the higher-order
terms on the right hand side (r.h.s.) of Eq.~\eqref{symb}
correspond to diagrams where, besides having the basic (quadratic in
$h_{\mu\nu}$) vertex $\mu_A$ on the $A$ worldline being connected by two
gravity propagators to two $GM_B$ ``sources'' on the $B$ worldline, we also
have some further gravity propagators connecting one of the worldlines either
to one of the worldline vertices, or to some intermediate ``field'' vertex.
Note that the information about the 1PN corrections to both gravito-electric 
($\mu_\ell$) and gravito-magnetic ($\sigma_\ell$) multipolar interactions 
(of any degree $\ell$) is contained in the work of 
Damour, Soffel and Xu~\cite{Damour:1991yw,Damour:1992qi,Damour:1993zn}.
We shall discuss below the effect of the subleading (post-Newtonian) terms 
in~\eqref{symb} on the EOB description of the dynamics of tidally interacting
binary systems.

%----------------------------------------------------
\section{Incorporating dynamical tidal effects in the 
Effective One-Body (EOB) formalism}
\label{sec:sec3}
%----------------------------------------------------

\subsection{General proposal}

The EOB formalism~\cite{Buonanno:1998gg,Buonanno:2000ef,Damour:2001tu}
 replaces the two-body interaction Lagrangian (or
Hamiltonian) by a Hamiltonian, of a specific form, which depends 
only on the relative position and momentum of the binary system, 
say $({\bm q},{\bm p})$. For a non spinning BBH system, it has been 
shown that its dynamics, up to the 3PN level, can be described by 
the following EOB Hamiltonian (in polar coordinates, within the 
plane of the motion):
\be
\label{eq:Heob}
H_{\rm EOB}(r,p_{r_*},p_\varphi) = M\sqrt{1+2\nu (\hat{H}_{\rm eff}-1)}
\ee
where
\be
\label{eq:Heff}
\hat{H}_{\rm eff} = \sqrt{p_{r_*}^2 + A(r) \left( 1 + \frac{p_{\varphi}^2}{r^2} + z_3 \, \frac{p_{r_*}^4}{r^2} \right)} \, .
\ee
Here $M=M_A + M_B$ is the total mass, $\nu \equiv M_A \, M_B / (M_A + M_B)^2$ is the symmetric 
mass ratio and $z_3 \equiv 2\nu (4-3\nu)$. In addition we are using rescaled dimensionless (effective) 
variables, notably $r = r_{AB} / GM$ and $p_{\varphi} = P_{\varphi} / (GM_A M_B)$, and $p_{r_*}$ is 
canonically conjugated to a ``tortoise'' modification of $r$~\cite{Damour:2009ic}.

A remarkable feature of the EOB formalism is that the complicated, 
original 3PN Hamiltonian (which contains many corrections to the basic 
Newtonian Hamiltonian $\frac{1}{2} \, {\bm p}^2 + 1/r$) can be replaced 
by the simple structure (\ref{eq:Heob}), (\ref{eq:Heff}) whose two crucial 
ingredients are: (i) a ``double square-root'' structure 
$H_{\rm EOB} \sim \sqrt{1+\sqrt{{\bm p}^2 + \cdots}}$, and (ii) the 
``condensation'' of most of the nonlinear relativistic gravitational 
interactions in one function of the (EOB) radial variable: 
the basic ``radial potential'' $A(r)$. In addition, the structure of the function 
$A(r)$ is quite simple. At the 3PN level it is simply equal to
\be
\label{eq:3.3}
A^{\rm 3PN} (r) = 1-2u+2 \, \nu \, u^3 + a_4 \, \nu \, u^4 \, ,
\ee
where $a_4 = 94/3 - (41/32)\pi^2$, and $u \equiv 1/r = GM/r_{AB}$. 
It was recently found~\cite{Damour:2009kr} 
that an excellent description of the dynamics of BBH systems 
is obtained by: (i) augmenting the presently computed terms 
in the PN expansion (\ref{eq:3.3}) by additional 
4PN and 5PN terms, and by (ii) Pad\'e-resumming the corresponding 
5PN ``Taylor'' expansion of the $A$ function. In other words, 
BBH (or ``point mass'') dynamics is well described by a function of the form
\be
\label{eq:3.4}
A^0(r) = P^1_5\left[1-2u+2\nu u^3 + a_4 \nu u^4 + a_5\nu u^5 + a_6\nu u^6\right]  ,
\ee
where $P^n_m$ denotes an $(n,m)$ Pad\'e approximant.
It was found in Ref.~\cite{Damour:2009kr} that a good agreement between
EOB and numerical relativity binary black hole waveforms is obtained 
in an extended ``banana-like'' region in the $(a_5,a_6)$ plane approximately 
extending between the points  $(a_5,a_6)=(0,-20)$ and $(a_5,a_6)=(-36,+520)$.
In this work we shall select the values $a_5=-6.37$, $a_6=+50$ which lie
within this good region.

Our proposal for incorporating dynamical tidal effects in the EOB formalism
consists in preserving the simple general structure
\eqref{eq:Heob}, \eqref{eq:Heff} of the EOB Hamiltonian, but to modify the BBH
radial potential~\eqref{eq:3.4} (which corresponds to the point-mass action~\eqref{eq:2.1n})
by augmenting it by some ``tidal contribution''. In other words the proposal
is to use Eqs.~\eqref{eq:Heob}, \eqref{eq:Heff} with
\be
\label{eq:3.5}
A(r) = A^0(r) + A^{\rm tidal} (r) \, .
\ee

\subsection{Incorporating leading order (LO) dynamical tidal interactions}

Let us show that, at the leading order (LO), one can use a tidal contribution of
the form
\be
\label{eq:3.6}
A^{\rm tidal}_{LO} (r) = -\sum_{\l\geq 2} \kappa_\ell^{\rm T} u^{2\ell + 2} , 
\ee
with some dimensionless coefficient $\kappa_\ell^{\rm T}$.

Indeed, if we keep only the Newtonian approximation of the full EOB Hamiltonian \eqref{eq:Heob}, 
\eqref{eq:Heff}  (using $A(r) \equiv 1 + \bar A (r)$ with $\bar A (r) = -2 \, GM / (c_0^2 \, r_{AB}) + \cdots$ 
being 1PN small as $1/c_0^2 \to 0$) one finds (with $\mu \equiv M^A M^B / M$)
\be
H_{\rm EOB} \simeq M \, c_0^2 +  \frac{1}{2} \mu\, {\bm p}^2 + \frac{1}{2} \mu \, \bar A (r) + 
{\mathcal O} \left( \frac{1}{c_0^2} \right) \, ,
\ee
which exhibits the role of $\frac{1}{2} \, \mu \, \bar A (r)$ as being the interaction energy. 
Decomposing $\bar A (r) = \bar A^0 (r) + A^{\rm tidal} (r)$, and remembering that there is a sign 
reversal between the interaction energy and the interaction Lagrangian, 
we see that the terms \eqref{eq:2.14} can be converted in a contribution to the $A(r)$ potential 
of the form \eqref{eq:3.6}, if the coefficients $\kappa_{\l}^{\rm T}$ take the values
\begin{eqnarray}
\label{eq:3.7}
\kappa_{\l}^{\rm T} &= &2 \, k_\l^A \, \frac{M_B}{M_A} \left( \frac{R_A \, c_0^2}{G (M_A + M_B)} 
\right)^{2\l + 1} \nonumber \\
&&+ \, 2 \, k_\l^B \, \frac{M_A}{M_B} \left( \frac{R_B \, c_0^2}{G(M_A + M_B)} \right)^{2\l + 1} 
\nonumber \\
&= &2 \, \frac{M_B \, M_A^{2\l}}{(M_A + M_B)^{2\l + 1}} \, \frac{k_\l^A}{c_A^{2\l + 1}} \nonumber \\
&&+ \, 2 \, \frac{M_A \, M_B^{2\l}}{(M_A + M_B)^{2\l + 1}} \, \frac{k_\l^B}{c_B^{2\l + 1}} \, .
\end{eqnarray}
In the second form, we have introduced the compactness parameters of the stars: 
$c_A \equiv GM_A / (R_A \, c_0^2)$. It is interesting to note that the dimensionless tidal parameters 
that enter the EOB dynamics are (when $M_A \sim M_B$) the ratios $k_\l^A / c_A^{2\l + 1}$, rather 
than the Love numbers $k_\l^A$. Let us also note that the velocity of light $c_0$ formally appears in 
the numerator of $\kappa_\l^{\rm T}$. This is related to the fact that, contrary to the coefficients of the 
successive powers of $u$ that enter the BBH EOB potential $A^0 (r)$ which are (roughly speaking) 
pure numbers of order unity, the coefficients $\kappa_\l^{\rm T}$ entering the tidal contribution 
$A^{\rm tidal} (r)$ will tend to be much larger than unity (and to increase with $\l$). For instance, we 
shall typically find that $\kappa_2^{\rm T} = {\mathcal O} (100)$. This numerical difference makes it 
consistent to add to $A^0 (r)$ (which is known for sure only up to $u^4$ terms, i.e. 
the 3PN level) additional terms $\propto u^6 + u^8 + \cdots$ that would formally correspond to 
5PN $+$ 7PN $+ \, \cdots$ contributions if their coefficients were ``of order unity'' 
(at least in the parametric sense).

 Finally, to illustrate the typical numerical values of the EOB tidal parameters we give 
in Table~\ref{tab:table1}  the values of $\kappa_2^{\rm T}$ for  three
paradigmatic systems, one equal-mass BNS and two BHNS of mass ratios $q\equiv M_{BH}/M_{NS}=4$ and $q=10$.
The neutron star model is described with a  ``realistic''EOS SLy (with a
piece-wise polytropic representation, see below) and has the following
characteristics: mass $M=1.35M_\odot$, compactness $c=0.17385$, radius $R=11.466$ km.
Note that the main dependence on the equation of state (EOS) in $\kappa^T_\ell$ (say for the equal-mass
BNS case) comes from $\kappa^T_\ell\propto R_A^{2\ell+1}$. Therefore, if one were
considering a NS of different radius (because of the use of a different EOS)
with the same mass, $\kappa^T_2$ would be approximately given by 
$\k^T_2\sim 73(R_A/11.466\,{\rm km})^5$

%========================================
% Table 1: tidal properties of BNS and BHNS systems
%========================================
\begin{table}[t]
\caption{\label{tab:table1}Tidal properties of BNS and BHNS system. The NS
 model is obtained using the piece-wise polytropic representation of EOS 
 SLy and has compactness $c = 0.17385$. Other properties of the model 
 can be found in Table~\ref{tab:table2}.}
 \begin{center}
  \begin{ruledtabular}
  \begin{tabular}{lcccc}
    Model &  $q$  & $\kappa^T_2$ &   $\k^{\rm T}_3$ &  $\k^{\rm T}_3$\\
   \hline \hline
     BNS   &  1    &  73.0426&  165.2966 & 509.6131       \\
     BHNS &  4   &    1.4959&   0.5416    & 0.2672        \\
     BHNS &  10 &    0.0726&  0.0054    &  0.0005       
  \end{tabular}
\end{ruledtabular}
\end{center}
\end{table}%
%===============================================

One sees in Table~\ref{tab:table1} that the dimensionless tidal
parameter $\kappa^T_2$ is a strongly decreasing function of the mass ratio.
This is analytically understood by looking at Eq.~\eqref{eq:3.7}. If the label $B$
refers to a black hole (so that $k_\ell^B=0$),
denoting $q\equiv M_{BH}/M_{NS}=M_B/M_A$, we have $\kappa^T_\ell=(\kappa^{\rm T}_\ell)^A$ 
where 
\be
(\kappa^{\rm T}_\ell)^A=2\dfrac{k^A_\ell}{c_A^{2\ell+1}}\dfrac{q}{(1+q)^{2\ell+1}}.
\ee
Here $c_A$ denotes as above the compactness of the NS.
Therefore, as soon as the mass ratio $q$ is significantly larger than one,
we see that $(\kappa^{\rm T}_\ell)^A$ contains a small factor
$q^{-{2\ell}}$ that suppresses the tidal contribution.
As a consequence, GW-observable tidal effects will be strongly suppressed 
in realistic BHNS systems.
Note, however, that it might be quite useful to compare numerical
relativity simulations of ``artificial'' BHNS systems of mass ratio $q\sim 1$
to their EOB description to probe the analytical understanding of the 
late inspiral and plunge phase. In particular, we note that, as a function of $q$, 
$\k_2^{\rm  T}\propto q/(1+q)^5$ vanishes both when $q\to 0$ and $q\to\infty$
and reaches a maximum value when $q=M_{BH}/M_{NS}=1/4$. Moreover the maximum
value of $\k_2^{\rm T}$ is larger than the value of $\k_2^T$ for a
corresponding {\it equal-mass} BNS system by a factor $4^6/5^5=1.311$.
We suggest that the numerical study of such astrophysically irrelevant
BHNS systems (with $M_{BH}/M_{NS}\sim 1/4$) can be quite useful for improving
our understanding of tidal interactions in strongly-interacting (near contact) 
regimes.

%------------------------------------------------------------------
\subsection{Parametrizing higher-order dynamical tidal corrections}
\label{sec:1pn_tides}
%------------------------------------------------------------------

Above we discussed the {\it leading order} (LO) contribution of
tidal interactions to the EOB ``radial potential'' $A(r)$.
We also discussed the structure of sub-leading (post-Newtonian) contributions
to tidal interactions, Eq.~\eqref{symb}. 
Comparing the structure~\eqref{symb} to the part of the EOB action linear in
$A^{\rm tidal}$, which is proportional to the product of $A^{\rm tidal}$ by
reduced mass $\mu=M^A M^B/(M^A+M^B)$, we see that the general
structure of the tidal contributions to the $A(r)$ potential is
\begin{align}
\label{eq:T2}
&A_{\mu_A}^{\rm tidal}\sim  \dfrac{M^A + M^B}{M^A M^B} \mu^A
\dfrac{(GM^B)^2}{r^{\ell +2}}\nonumber\\
&\times\left[1 + \dfrac{GM^A}{r} + \dfrac{GM^B}{r} 
+ \left(\dfrac{GM^A}{r}+\dfrac{GM^B}{r}\right)^2 + \dots\right] 
\end{align}
where we invoked dimensional analysis to insert appropriate powers
of the (EOB) radial separation $r$.
[Contrary to the action~\eqref{symb} which also depends on velocities (and higher-derivatives),
the EOB radial potential depends only on the radius $r$.]

In other words, if we separate, for each multipolar order, the $\mu_A$ and
$\mu_B$ contributions to $A^{\rm tidal}$,
\be
\label{eq:T3}
A^{\rm tidal} = \sum_{\ell \geq 2} A^{\mu^A_\ell} + \sum_{\ell \geq 2}
A^{\mu_\ell^B} ,
\ee
we can write
\begin{align}
\label{eq:T4}
A^{\mu^A_\ell} = A^{\mu_\ell^A}_{\rm LO}\left[1 + \alpha_1^{A(\ell)} u +
  \alpha_2^{A(\ell)}u^2+\alpha_3^{A(\ell)}u^3 \dots\right],
\end{align}
where
\be
\label{eq:T5}
A^{\mu_\ell^A}_{\rm LO} \equiv - \kappa_\ell^A u^{2\ell +2}
\ee
is the part of $A^{\rm tidal}_{\rm LO}$, Eq.~\eqref{eq:3.6},
which is linear in $\mu^A_\ell$, or $k_\ell ^A$, i.e.
\be
\kappa^A_\ell = 2 \, k_\l^A \, \frac{M_B}{M_A} \left( \frac{R_A \, c_0^2}{G (M_A + M_B)} 
\right)^{2\l + 1}.
\ee
Similarly, one will have 
\be
\label{eq:T7}
A^{\mu_\ell^B} = A_{\rm LO}^{\mu_\ell^B} \left[1 + \alpha_1^{B(\ell)} u 
+\alpha_2^{B(\ell)} u^2 + \alpha_2^{B(\ell)} u ^3 + \dots \right]
\ee
The coefficient $\alpha_1^{A(\ell)}$ represents the next to leading order (NLO)
fractional correction to the leading order $A^{\mu_\ell^A}_{\rm LO}$ (i.e. a
1PN fractional correction), while $\alpha_2^{A(\ell)}$ represents the
next-to-next to leading order (NNLO) correction (i.e. a 2PN fractional
correction), etc.
These coefficients are not pure numbers, but rather function of the two
dimensionless mass ratios
\begin{align}
\label{eq:T8}
X_A &\equiv \dfrac{M_A}{M_A+M_B},\\
X_B &\equiv \dfrac{M_B}{M_A+M_B}\equiv 1 - X_A .
\end{align}
The coefficients entering Eq.~\eqref{eq:T7} are obtained from those
entering~\eqref{eq:T4} by the interchange of $X_A$ and $X_B$, i.e.
$\alpha^{A(\ell)}_n (X_A,X_B)=\alpha^{B(\ell)}_n (X_B,X_A)$. The
symbolic structure~\eqref{eq:T2} would naively suggest that
$\alpha_1^{A(\ell)}$ is a linear combination of $X_A$ and $X_B$ and
that $\alpha_2^{A(\ell)}$ is a combination of $X_A^2$, $X_AX_B$ and $X_B^2$.
However, as the reformulation of~\eqref{symb} in terms of an EOB potential
\eqref{eq:T2} involves a ``contact transformation'' that depends on the
symmetric mass ratio $\nu\equiv X_AX_B$ (see Ref.~\cite{Buonanno:1998gg}), the mass-ratio
dependence of $\alpha_n^{A(\ell)}$ might be more complicated. 
Note that, by using the identity $X_A+X_B\equiv 1$, one can, e.g., express
$\alpha_n^{A(\ell)}$ in terms of $X_A$ only.
[Then  $\alpha_n^{B(\ell)}$ will be the same function of $X_B$ than
$\alpha_n^{A(\ell)}$ of $X_A$.]
Note also that, if one wishes, one can, for each value of $\ell$ factorize
the total LO terms  $-\kappa_\ell^{\rm T} u^{2\ell +2}$, and write
\begin{align}
\label{eq:T9}
A^{\rm tidal}=\sum_{\ell\geq 2} - \kappa_\ell^{\rm T}u^{2\ell+2}\hat{A}^{\rm tidal}_\ell,
\end{align}
where
\be
\label{eq:T10}
\hat{A}^{\rm tidal}_\ell\equiv 1
  +\bar{\alpha}_1^{(\ell)}u + \bar{\alpha}_2^{(\ell)}u^2 + \dots ,
\ee
with 
\be
\label{eq:T11}
 \bar{\alpha}_n^{(\ell)}\equiv \dfrac{\k_\ell^A \alpha_n^{A(\ell)} + \k_\ell^B\alpha_n^{B(\ell)}  }{\k^A_\ell + \k_\ell^B}.
\ee
Using Eqs.~(4.27) and (4.29) of~\cite{Damour:1992qi}, or Eq.~(3.33) of~\cite{Damour:1993zn},
together with effective action techniques, a recent calculation ~\cite{DEF09}
gave the following result for the 1PN coefficient of multipolar order
$\ell=2$, $\alpha_1^{A(2)}$, namely
\be
\label{eq:alpha1}
\alpha_1^{A(2)}=\dfrac{5}{2}X_A.
\ee 
More work is needed to determine the higher degree and/or higher order 
 coefficients $\alpha_n^{A(\ell)}(X_A,X_B)$,
and thereby the coefficients $\bar{\alpha}_n^{(\ell)}$ entering Eq.~\eqref{eq:T11}.
Below, we shall focus on the equal-mass case where the coefficients  $\alpha_n^{A(\ell)}$
become pure numbers.

Here we shall explore three possible proposals for including higher-order
PN corrections in tidal effects. The first proposal consists in truncating
Eq.~\eqref{eq:T10} at 1PN order in a straightforward ``Taylor'' way, i.e.
to consider a PN correcting factor to the EOB radial potential of the form
\begin{equation}
\label{eq:linear}
\hat{A}^{\rm tidal}_\ell = 1+\bar{\alpha}_1^{(\ell)} u. 
\end{equation}
The second proposal consists in considering a PN correcting factor which
has a ``Pad\'e-resummed'' structure, i.e.
\begin{equation}
\label{eq:pade}
\hat{A}^{\rm tidal}_\ell = \left( 1- \bar{\alpha}_1^{(\ell)} u\right)^{-1} .
\end{equation}
Our third proposal consists in considering a PN correcting factor which would
result from having a ``shift'' between the EOB radial coordinate and the radial 
coordinate appearing most naturally in a Newtonian-like tidal interaction
($\propto 1/r^{2\ell+2}$).
\begin{equation}
\label{eq:harmonic}
\hat{A}^{\rm tidal}_\ell = \left( 1- \widetilde{\alpha}_1^{(\ell)} u\right)^{-(2\ell +2)}.
\end{equation}
We use here a different notation  for the 1PN coefficient, 
$\widetilde{\alpha}_1^{(\ell)}$, as a reminder that, for instance, 
when  $\ell=2$, the parametrization~\eqref{eq:harmonic} corresponds
to a 1PN coefficient in the parametrization~\eqref{eq:linear} given
by 
\be
\label{alpha_tilde}
\bar{\alpha}_1^{(2)}=6\,\widetilde{\alpha}_1^{(2)}.
\ee

%--------------------------------------------------------------------------------
\section{Comparing EOB to numerical relativity results on "waveless" circular binaries   }
\label{sec:nr}
%--------------------------------------------------------------------------------
The aim of this section is to compare stationary quasi-circular
configurations of neutron star binaries computed, on the one hand, 
in the analytical framework outlined above and, on the other hand,
in the numerical framework recently implemented by 
Ury${\rm\bar{u}}$ et al.~\cite{Uryu:2009ye} (see also~\cite{Uryu:2005vv}).
The quantity from both frameworks that we shall compare is the
binding energy $E_b$ as a function of the orbital frequency $\Omega$.

\subsection{Tidally interacting BNS circular configurations in the EOB framework }
\label{ar:circular}

\subsubsection{BNS binding energy in the EOB framework}
\label{sbsc:eob_circular}
As an application of the formalism discussed so far, 
we consider in this section binaries in exactly circular orbits,
in absence of radiative effects (these will be discussed in the following section).

As the EOB formalism is based on a Hamiltonian 
description of the conservative dynamics, the stable circular orbits correspond to minima, with respect 
to $r$, of the radial potential $H_{\rm EOB}^{\rm radial} (r,p_{\varphi}) \equiv H_{\rm EOB} (r , 
p_{r_*} = 0, p_{\varphi})$. Minimizing $H_{\rm EOB}^{\rm radial} (r,p_{\varphi})$ is equivalent to 
minimizing the corresponding effective Hamiltonian $\hat H_{\rm eff}$, or, its square, i.e.
\begin{eqnarray}
\label{eq:5.1}
(\hat H_{\rm eff}^{\rm radial})^2 \, (r,p_{\varphi}) &= &A(r) \left( 1 + \frac{p_{\varphi}^2}{r^2} \right) 
\nonumber \\
&\equiv &A(u) + p_{\varphi}^2 \, B(u) \, .
\end{eqnarray}
Here, we have used the short-hand notation $u \equiv 1/r = GM/R$ and $B(u) \equiv u^2 \, A(u)$. 
Minimizing (\ref{eq:5.1}) with respect to $r$ (or, equivalently, $u$), for a given (scaled) total angular 
momentum $p_{\varphi} \equiv J^{\rm tot} / GM\mu$, yields the following equation
\be
\label{eq:5.2}
A'(u) + p_{\varphi}^2 \, B'(u) = 0,
\ee
where the prime denotes a $u$-derivative. This leads to the following
parametric representation of the squared angular momentum:
\begin{equation}
j^2(u)=-\dfrac{A'(u)}{(u^2 A(u))'}\quad\text{(circular orbits)},
\end{equation}
where we use the letter $j$ to denote the value of $p_\varphi$ along
the sequence of circular orbits.
Inserting this $u$-parametric representation of 
$j^2$ in Eq.~\eqref{eq:Heff} defines the $u$-parametric representation of the
effective Hamiltonian $\hat{H}_{\rm eff}(u)$.
We can then obtain (at least numerically) $\hat{H}_{\rm eff}$ as a function of $x$
by eliminating $u$ between $\hat{H}_{\rm eff}(u)$ and the corresponding
$u$-parametric representation of the frequency parameter $x=(GM\Omega/c^3)^{2/3}$
obtained by the angular Hamilton equation of motion in the circular case
\begin{equation}
\label{eq:Omega}
M\Omega(u) = \dfrac{1}{\mu}\dfrac{\de H_{\rm EOB}}{\de j}=\dfrac{M A(u)j(u) u^2}{H_{\rm real}\hat{H}_{\rm eff}},
\end{equation}
where $H_{\rm real}$ denotes the real EOB Hamiltonian
\begin{equation}
\label{eq:real_hamiltonian}
H_{\rm EOB} = M \sqrt{ 1 + 2\nu\left( \hat{H}_{\rm eff} - 1\right)}.
\end{equation}
In this situation, the binding energy $E_b$ of the system is simply given
by
\be
\label{eq:Eb}
E_b(\Omega) = H_{\rm EOB}-M = M\left\{ \sqrt{ 1 + 2\nu\left( \hat{H}_{\rm eff} - 1\right)}-1 \right\},
\ee
where $M$ denotes, as above, the total mass $M=M_A+M_B$ of the system,
and where one must eliminate $u$ between Eq.~\eqref{eq:Omega} and
Eq.~\eqref{eq:Eb} to express the r.h.s. in terms of $\Omega$. Note that
the function $E_b(\Omega)$ depends also on the choice of the following
parameters: $\kappa_\ell^T$, $\alpha_1^{A(\ell)}$ and $\alpha_1^{B(\ell)}$.
Here we shall focus on the equal-mass case, and consider the dependence
of $E_b(\Omega)$ only on $(\kappa_2^T,\kappa_3^T,\kappa_4^T)$ and restrict
the parametrization of 1PN tidal effects to the consideration of a 
{\it single} 1PN tidal parameter  $\bar{\alpha}_1$ that is taken to
be the same for the three values of $\ell$ that we consider.
In addition, we will incorporate 1PN corrections to tidal effects in
the three aforementioned functional forms, 
Eq.~\eqref{eq:linear}-\eqref{eq:harmonic} and contrast their performances.

%-----------------------------------------------------
\subsubsection{BNS binding energy in the PN framework}
\label{sbsc:pn_circular}
%----------------------------------------------------
We also want to constrast the performance of the EOB approach 
(which represents a resummation of the dynamics of the binary system) 
with the ``standard'' nonresummed PN-based description of the 
binding energy of tidally interacting BNS, as used for instance in
Ref.~\cite{Mora:2003wt}. 
The PN-expanded binding energy is written in the form
\be
E_b(\Omega) = E_{\rm point-mass}(\Omega) + E^{\rm tidal}(\Omega),
\ee
where
\begin{align}
&E_{\rm point-mass}(\Omega) =-\dfrac{\mu}{2} x\bigg\{1-\left(\dfrac{3}{4}+\dfrac{1}{12}\nu\right)x \nonumber\\
&-\left(\dfrac{27}{8}-\dfrac{19}{8}\nu+\dfrac{1}{24}\nu^2\right)x^2\nonumber\\
&-\left(\dfrac{675}{64}-\left[\dfrac{34445}{576}-\dfrac{205}{96}\pi^2\right]\nu+\dfrac{155}{96}\nu^2 
+ \dfrac{35}{5184}\nu^3\right)x^3\bigg\},
\end{align}
is the 3PN accurate post-Newtonian binding energy of two point-masses as
function of the orbital frequency parameter $x=(GM\Omega/c^3)^{2/3}$ 
~\cite{Damour:1999cr,Damour:2001bu}. The expression of the tidal contribution
$E^{\rm tidal}(\Omega)$ can be obtained for all values of the multipolar 
index $\ell$ by noting the following. Any (perturbative) power-law radial contribution 
to the interaction Hamiltonian of the form
\be
\delta H(r) = -\dfrac{c_n}{r^n}
\ee
is easily shown to contribute a corresponding term
\be
\delta E_b(\Omega) = +\left(\dfrac{2}{3}n-1\right)\dfrac{c_n}{r_\Omega^n},
\ee  
where it should be noted that the sign of the tidal contribution flips
between the Hamiltonian and the binding energy expressed as a function
of the orbital frequency ($r_\Omega$ denoting the Newtonian  value of $r$
corresponding to a given circular orbit of frequency $\Omega$). As
a result, we have the leading order contribution to the PN-tidal
contribution
\be
\label{eq:dEtidal}
E^{\rm tidal}_{\rm LO}(\Omega)= + \dfrac{\mu}{2}\sum_{\ell\geq 2} \left[\dfrac{2}{3}(2\ell+2)-1\right]\kappa_\ell^T\, x^{2\ell +2}.
\ee
We shall also explore the effect of correcting $E^{\rm tidal}_{\rm LO}$
by a fractional 1PN contribution, i.e. to employ a PN tidal contribution 
of the form
\be
\label{eq:NLO_pn}
E^{\rm tidal}(x)= (1+ \bar{\alpha}_1'x)E^{\rm tidal}_{\rm LO}(x).
\ee
where the (approximate) link with the previously defined $\bar{\alpha}_1$ is
\be
\bar{\alpha}_1'=\dfrac{11}{9} \bar{\alpha}_1.
\ee
Here the numerical coefficient $11/9$ arises as a consequence of the factor $2n/3-1$ in
the result above (considered for $n=6$ and $n=7$).

%--------------------------------------------------------------------
\subsection{BNS circular configurations in numerical relativity}
\label{nr:circular}
%-------------------------------------------------------------------
%=====================================
% TABLE 2:  Properties of BNS model considered
%=====================================
\begin{table*}[t]
\caption{\label{tab:table2} Properties of NS models considered discussed in the numerical analysis of Ref.~\cite{Uryu:2009ye}. The EOS are
represented as piece-wise-polytropic functions (on four intervals) as proposed in~\cite{Read:2008iy,Read:2009yp}. For the models considered, the present
table is compatible with Table III of ~\cite{Uryu:2009ye}. From left to right, the columns report: the dividing density between the low-density
part (the crust) and the higher density part of the EOS; the four adiabatic indices for each polytropic interval, $\{\Gamma_0,\Gamma_1,\Gamma_2,\Gamma_3\}$; 
the compactness $c=M/R$; the NS mass $M$ and the NS radius $R$; the Love numbers $k_2$, $k_3$ and $k_4$.}
\begin{center}
  \begin{ruledtabular}
  \begin{tabular}{lccccccccccc}
    Model   &  $\log(\rho_0)$
            & $\Gamma_0$
            & $\Gamma_1$ 
            & $\Gamma_2$
            & $\Gamma_3$
            & $M/R$ 
            & $M$ 
            & $R$
            & $k_2$  
            & $k_3$
            & $k_4$\\
    \hline \hline
 2H    &13.847 &1.35692&3    &3    & 3    &0.13097&1.3507& 15.229 & 0.1342& 0.0407& 0.0168   \\
 HB    &14.151 &1.35692&3    &3    & 3    &0.17181&1.3507& 11.608 & 0.0946& 0.0260&0.0097  \\  
 2B    &14.334 &1.35692&3    &3    & 3    &0.20500&1.3505&  9.728 & 0.0686& 0.0174& 0.0059  \\
 SLy   &14.165 &1.35692&3.005&2.988& 2.851&0.17385&1.3499& 11.466 & 0.0928& 0.0254& 0.0095\\   
 FPS   &14.220 &1.35692&2.985&2.863& 2.600&0.18631&1.3511& 10.709 & 0.0805&0.0214&0.0077 \\
BGN1H1 &14.110 &1.35692&3.258&1.472& 2.464&0.15792&1.3490& 12.614 & 0.1059& 0.0307&0.0120 \\
  \end{tabular}
\end{ruledtabular}
\end{center}
\end{table*}
%========================================================================

\subsubsection{Numerical framework of Ury${\bar{u}}$ et al.}
\label{framework}

In a recent paper, Ury${\rm\bar{u}}$ et al.~\cite{Uryu:2009ye} constructed BNS
systems in quasi-circular orbits by solving numerically the full
set of Einstein's equations. The important advance of this work
with respect to previous analyses is the fact that Einstein equations
are solved for all metric components, including the nonconformally
flat part of the spatial metric. This goes beyond the common
{\it conformally flat}  
approximation that is usually employed for the spatial geometry.
The conformally flat approximation introduced systematic
errors which enter the PN expansion already at the 2PN level 
[see the detailed calculation in the Appendix B of Ref.~\cite{Damour:2000we}].
Consistently with this analytical argument, it was found
in Ref.~\cite{Uryu:2009ye} that the difference between conformally
flat and nonconformally flat calculations is so large that it can
mask the effect of tidal interaction for close systems. See, in
this respect, the location of the conformally flat (IWM) binding energy
curves in the two upper panels of Fig.~3 in Ref.~\cite{Uryu:2009ye}.
Below we shall however emphasize that the nonconformally flat calculations
of~\cite{Uryu:2009ye} still introduce significant systematic errors which enter
the PN expansion {\it at the 3PN level}. 

Since the new nonconformally flat results of Ury${\rm\bar{u}}$ et al. represent
a definitive improvement with respect to previous calculations, 
it is appealing to see to what extent these new results agree 
with existing analytical descriptions.
We extracted from Ref.~\cite{Uryu:2009ye} the six models which
present the highest computational accuracy. These models  were obtained
by using EOS labelled 2H,HB,2B, SLy, FPS and BGN1H1. These labels refer
to piecewise polytropic EOS. Note that in the case of SLy, FPS and BGN1H1
the corresponding piecewise polytropic EOS were proposed in
Ref.~\cite{Read:2008iy} as approximations to original tabulated EOS.
In the case of FPS and SLy, this implies that
the tidal coefficients $k_\ell$ that we have computed for this work 
differ (by $\sim 20\%$) from the ones that we had previously
computed in Ref.~\cite{Damour:2009vw} that used the original tabulated EOS.
For example, in the case of a neutron star model described by the 
SLy EOS and having a compactness $c=0.176$ (which corresponds to
a mass of $1.4M_\odot$), we obtain a dimensionless Love number
$k_2^{(\rm tab)}=0.07699$ (which is consistent with the first line of Table~I
of Ref.~\cite{Hinderer:2009}) if we use the tabulated EOS, while
we obtain $k_2^{(\rm ppoly)}=0.09123$ if we use the piece-wise
polytropic EOS. Note that the piece-wise polytropic result is $18.5\%$
larger than the tabulated one. This suggests that one should
refine the piece-wise polytropic approximation to realistic
tabulated EOS by incorporating $k_2$ within the set of observables
that are fitted.

Among the six EOS that we retain, three, i.e. 2H, HB and 2B, 
use two polytropic intervals, 
while the other three, i.e. SLy, FPS and BGN1H1,  
use four polytropic intervals.
We will thus have one dividing density\footnote{Here, following the notation
of~\cite{Read:2008iy}, we use the letter $\rho$ to denote the rest-mass
(baryon) density which was denoted by $\mu$ in our previous
work~\cite{Damour:2009vw}}, denoted by $\rho_0$, for 2H, HB and 2B,
and three dividing densities, $(\rho_0,\rho_1,\rho_2)$, for SLy, FPS and BGN1H1.
Here, $\rho_0$ indicates the dividing density between the lower density 
interval that approximates the subnuclear density part of the EOS 
(the crust) and the supernuclear density part.
The values of (the base-ten logarithm of) $\rho_0$ are displayed in 
the first column of Table~\ref{tab:table1}. 
For all EOS, the lower density interval (``crust'') is 
approximated by setting $(\Gamma_0,K_0)=(1.35692,3.59389\times 10^{13})$,
where $K_0$ (here is in cgs units)  gives the pressure $p$ in dyn/cm$^2$.
The other dividing densities (for the four-parameter EOS) are fixed 
as $\rho_1=10^{14.7}$ and $\rho_2=10^{15}$.
The corresponding adiabatic indices, $\{\Gamma_1,\Gamma_2,\Gamma_3\}$,
taken from~\cite{Read:2008iy,Uryu:2009ye} are also given in  Table~\ref{tab:table1}.
For the implementation of the piecewise polytropic EOS we follow
the procedure explained in Sec.~III of~\cite{Read:2008iy} and in 
Sec.~IID of~\cite{Uryu:2009ye}.

For each selected EOS, we computed the sequence of equilibrium models with the
 related Love numbers $k_\ell$ up to $\ell=4$. For the compactnesses corresponding
 to those used in~\cite{Uryu:2009ye} we display in Table~\ref{tab:table2}  the $k_\ell$'s
 together with the values of  mass and radius that we obtained from our calculation,
 to check consistency with the corresponding values of Table~III of ~\cite{Uryu:2009ye}.
 The small differences (at the $10^{-3}$) level are probably due to the fact that we 
 use the finite-digit value of the dividing density $\rho_0$ that they published.

\subsubsection{Subtracting tidal effects from NR data}

Let us start by noting two facts, that can be  checked from
the analytical expressions above, about the dependence of the binding
energy on the tidal parameters $\kappa_\ell^T$: i) this dependence is
to a very good approximation {\it linear} and ii) the numerical effect
of the $\kappa_2^T$ strongly dominates over that of the higher degree
$\k_\ell^T$'s. For example, if we take the tidal coefficients listed
in Table~\ref{tab:table1} (which correspond to the SLy EOS, which yields
a radius $\sim 11.5$ km for $1.35M_\odot$, which is in the middle of the
realistic range of NS radii) we find that the tidal contributions to
the binding energy would reach, if  they were extended to the maximum
frequency that we shall explore here, namely $M\Omega_{\rm max}=0.060$,
the following values: the $\kappa_2^T$ contribution to $E_b/M$ 
is $\sim 3.6\times 10^{-4}$; the $\kappa_3^T$ contribution 
is smaller than the $\kappa_2^T$ by a factor 0.053, and the
$\kappa_4^T$ is smaller than the  
$\kappa_2^T$ one by a factor $\sim3.85\times 10^{-3}$.

These two facts allow us to {\it approximately subtract tidal effects 
from NR data}. Indeed, if we assume that the binding energy computed
with a certain equation state $(EOS)_I$ is approximately given by
\be
E_b(\Omega;\, I) \approx h_0(\Omega) + (\kappa_2^T)_I h_2(\Omega)
\ee 
we can use the NR data for two different EOS, labelled by $(I,J)$ 
to compute, {\it  separately}
\begin{align}
\label{eq:h0}
h_0(\Omega) &\approx \dfrac{(\kappa_2^T)_I E_b(J)-  (\kappa_2^T)_J E_b(I)  }
                          {(\kappa_2^T)_I -  (\kappa_2^T)_J},\\
\label{eq:h2}
h_2(\Omega) &\approx \dfrac{E_b(I)-   E_b(J)  }
                          {(\kappa_2^T)_I -  (\kappa_2^T)_J} .
\end{align}
%------------------------
% FIG.1 h0 analysis
%------------------------
\begin{figure*}[t]
\begin{center}
\includegraphics[width=85 mm, height=70mm]{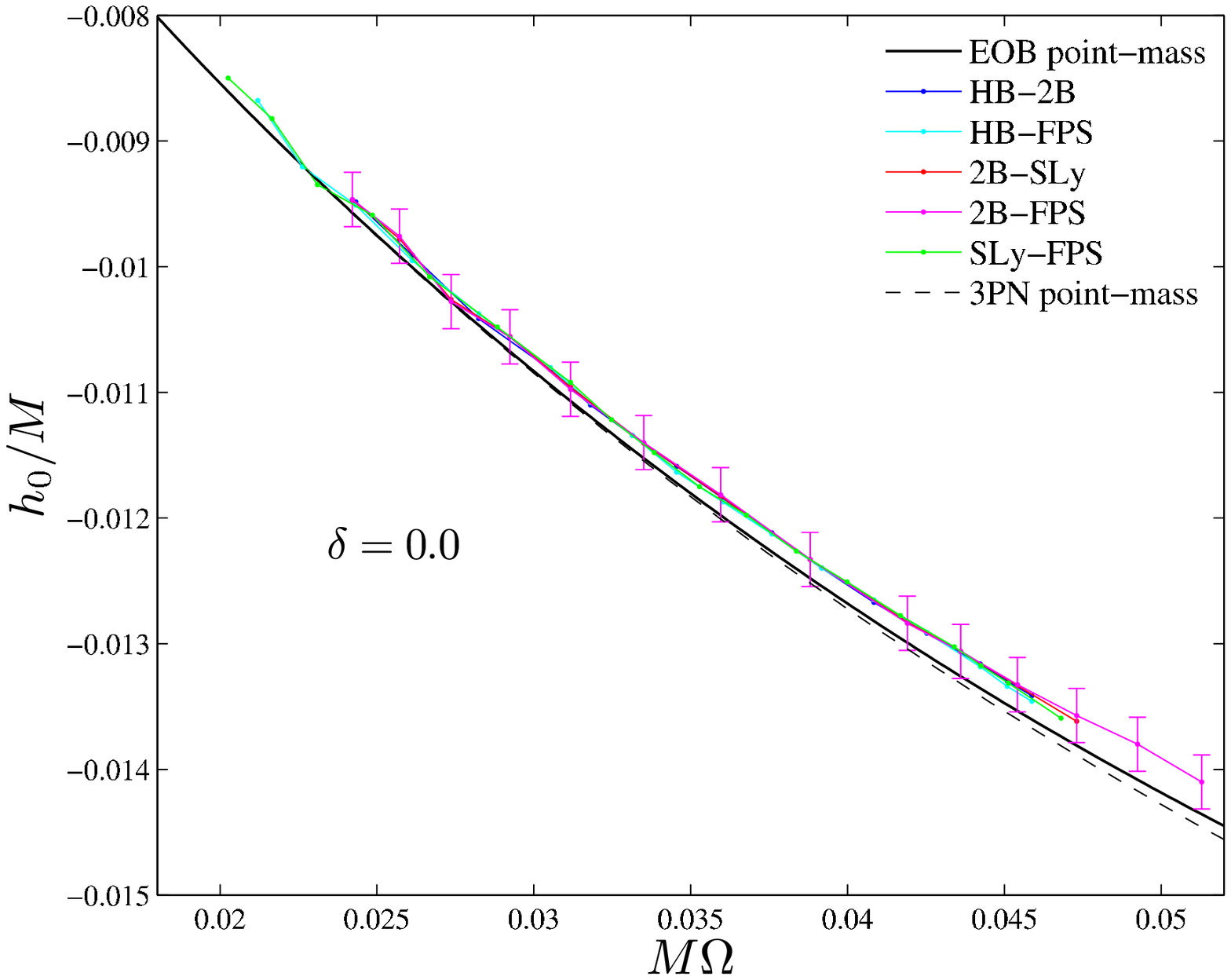}\hspace{5 mm}
\includegraphics[width=85 mm, height=70mm]{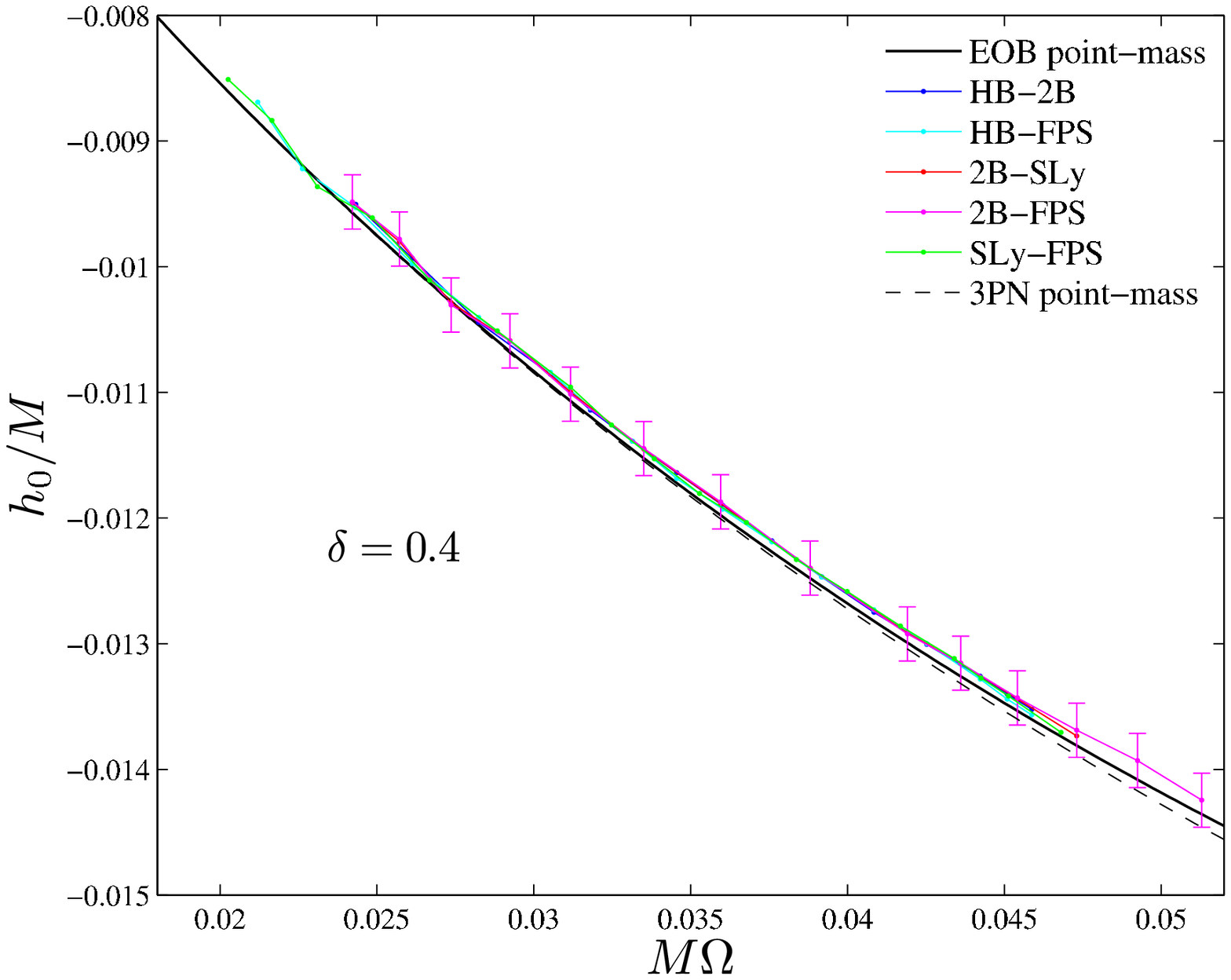}\\
\vspace{5mm}
\includegraphics[width=85 mm, height=70mm]{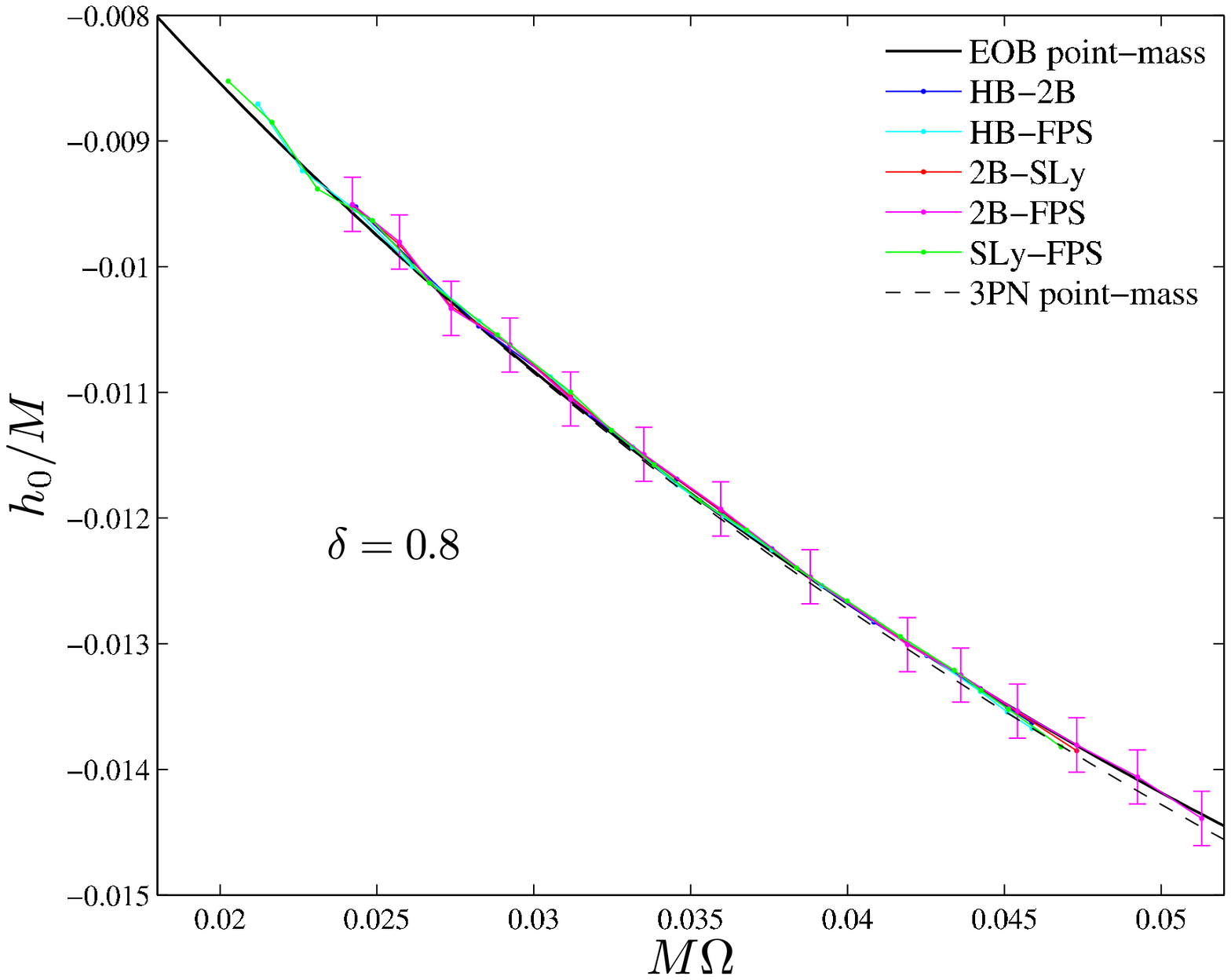}\hspace{5 mm}
\includegraphics[width=85 mm, height=70mm]{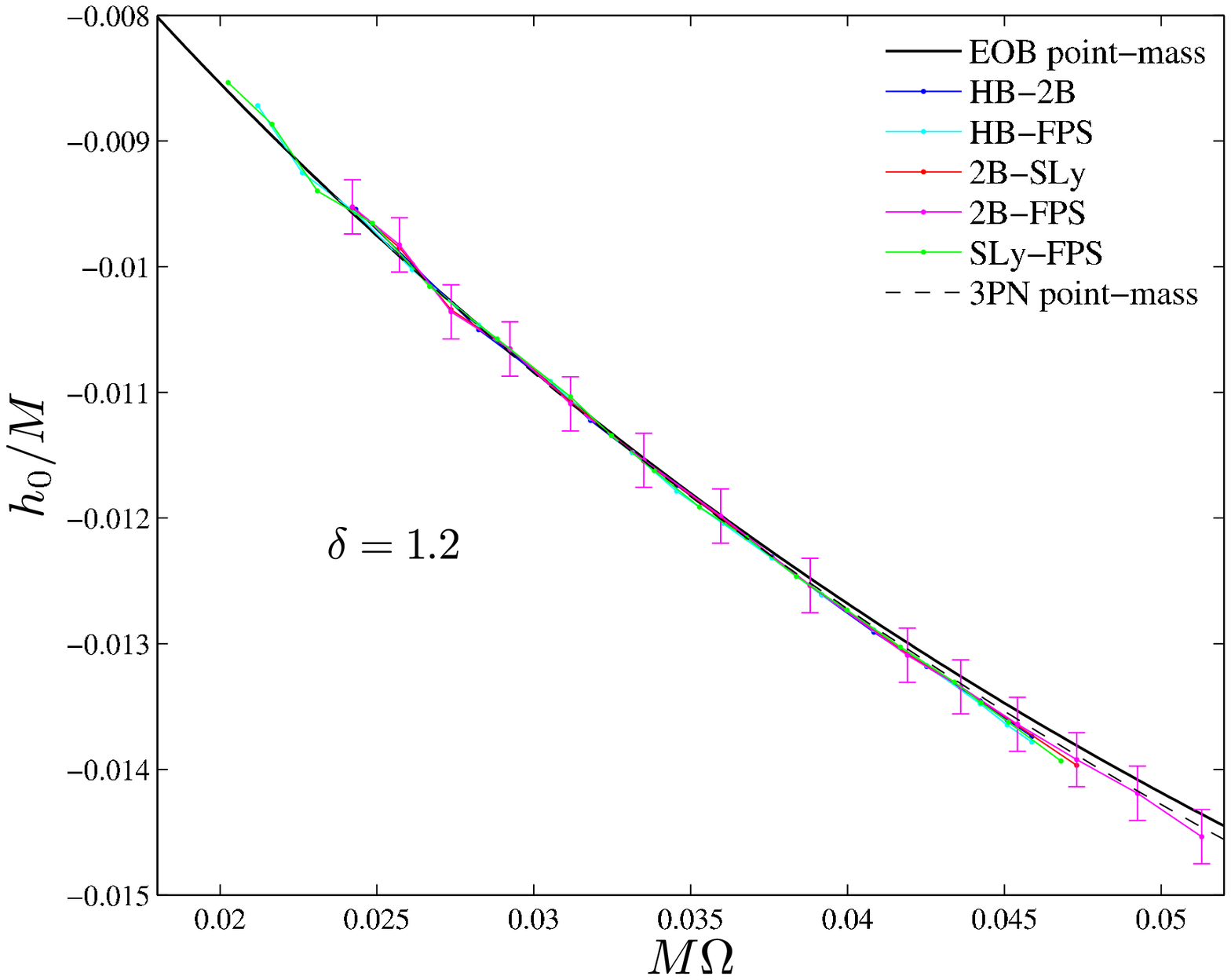}\\
\caption{\label{fig:fig1} Comparison between various $\delta$-corrected $h_0$'s (defined in Eq.~\eqref{eq:h0}) 
and the EOB (resummed, solid line) and 3PN (nonresummed, dashed line) point-mass representations of the binding energy. }
  \end{center}
\end{figure*}
%------------------------

Most importantly we see that Eq.~\eqref{eq:h0} allows us to compute
from the binding energies of two BNS sequences a third binding energy
function, $h_0$, which approximately represents the binding energy
of {\it non tidally interacting} neutron stars, i.e. the binding 
energy curve of two point-masses. 
The result of computing the r.h.s. of Eq.~\eqref{eq:h0} for five pairs $(I,J)$
of EOS having sufficiently different $\k^T_2$'s   is displayed in
the left panel of Fig.~\ref{fig:fig1}. Two important lessons can be
drawn from this figure: i) The subtraction procedure defined by
Eq.~\eqref{eq:h0} is remarkably able to define ``tidal-free'' energy
curves that are essentially on top of each other; this confirms 
that our procedure succeeds in subtracting out the EOS-dependence
of the binding energy curves; ii) However, the resulting ``universal''
$h_0$ curve still differs significantly both from the EOB point-mass
curve (black solid line) and the PN point-mass one (black dashed line).
This second issue will be addressed in the next subsection.

We shall not display here the result of computing the $h_2$ part
of the binding energy curve, Eq.~\eqref{eq:h2}, because it is more
sensitive than $h_0$ both to numerical noise (in the original NR data)
and to the presence of higher-order tidal PN contributions.
Below, we shall address the issue of determining the tidal contributions to
$E_b$  with a different approach.

\subsubsection{Detecting and subtracting systematic errors in NR data}
Here we address the issue ii) mentioned in the previous subsection.
Indeed, our subtraction procedure has given us access to the ``universal'',
EOS-independent part of the energy curve $h_0$. 
However, we have seen that $h_0$ still significantly differs from the
analytical point-mass models. We think that the origin of this discrepancy
is the presence of remaining ``systematic errors'' in the current
nonconformally flat approach to BNS systems. Though the nonconformally flat
integration scheme of Ury${\rm\bar{u}}$ et al. is an improvement over previous work,
it is however still only an approximation to the exact solution describing
two BNS interacting in a (conservative) ``time-symmetric'' 
manner (half-retarded-half-advanced). Here we shall only use the data obtained
by Ref.~\cite{Uryu:2009ye} called the ``waveless'' approximation.
In their approach, ``waveless'' means setting to zero the time-derivative
of the conformal spatial metric (in a certain gauge): $\de_t\tilde{\gamma}_{ab}=0$.
As the NR gauge is rather similar to the ADM-TT gauge used in the 3PN
calculation of the interaction Hamiltonian of a two point-mass 
system in Refs.~\cite{Jaranowski:1997ky,Damour:2001bu}, we can see,
by looking at the analytical expression of the 3PN-accurate ADM Hamiltonian,
that neglecting the terms containing $\pi_{ab}^{\rm TT}\sim \de_t\tilde{\gamma}_{ab}$
means neglecting some of the terms that contribute at the {\it 3PN level}.
[The simplest of these terms being the ``kinetic energy'' term proportional
to $\int d^3 x (\pi_{ab}^{\rm TT})^2$].
This analytical argument suggests that the current NR data miss some 3PN
contributions, i.e. they miss some terms proportional to $x^4$ in the binding
energy curve. We are therefore entitled in assuming that the discrepancy
displayed in the left-panel of Fig.~\ref{fig:fig1} between the NR $h_0$
and the point-mass analytical curves is, to leading order, given by
an expression of the type $\Delta E_b(\Omega) = \delta  \, x^4$ with an EOS-independent 
numerical coefficient $\delta$ that we expect to be of order unity.
Indeed, the right panel of Fig.~\eqref{fig:fig1} exhibits the fact that,
by subtracting $\Delta E_b(\Omega) = \delta \, x^4$, with $\delta= 0.8$ 
(see below) from all the individual  $h_0$ curves, we can reach a good 
visual agreement with both analytical point-mass models.
[Note that the approximate ``best-fit'' value of $\delta$  is mainly determined 
by the discrepancy 
NR/AR on the lower frequency part of the panel, say for $M\Omega< 0.035$
where the contribution to tidal effects is relatively negligible]. 

The remaining differences in this right panel are 
compatible with the known level of numerical errors in the NR 
data (see Fig.~4 of Ref.~\cite{Uryu:2009ye}).
Indeed~\cite{Uryu:2009ye} has used the virial theorem to gauge 
some of the systematic errors in their calculation by comparing 
two measures of the total mass of the system (Komar and ADM). 
The resulting (absolute value) differences
in binding energy, say $\delta^v E_b$ are in general at the 
level $10^{-4}M$. We used these differences to estimate formal
``error-bars'' on the  various energy curves that we use in this
work.
More precisely, in $E_b$ energy curves we add error bars of one-sided amplitude
$\pm \frac{1}{2}\delta E_b$, so that the length of the two-sided error
bars corresponds to the ``virial'' error.
As Fig.~\ref{fig:fig1} concerns a quantity, $h_0$, defined as a 
linear combination of NR data (see Eq.~\eqref{eq:h0}), we conservatively
estimated error bars on the $h_0$ curve corresponding to the pair 2B-FPS
by linearly combining
in absolute values the corresponding individual errors. We use this error
bar to gauge the quality of the other $h_0$ curves (which do not extend as
far in the high frequency range).
This conservative
estimate of the total error seems appropriate to the present situation
where the errors are not random, but rather systematic.
[Note, however, that these ``error-bars'' seem to be too conservative in
the lower frequency part of the panels because they exceed the ``distance''
between the $h_0$ curves and the point-mass models.]
Using these error bars we can now roughly estimate a range of  acceptable 
 values  of the NR correcting parameter $\delta$. As illustrated in the
four panels of Fig.~\ref{fig:fig1}, the range $0.4\leq \delta\leq 1.2$
is such that the $\delta$-corrected NR-deduced $h_0$ curves are within 
``one formal sigma'' from both point-mass analytical models.
We shall use this range below to estimate a corresponding range of 
probable values of the 1PN tidal parameter $\bar{\alpha}_1^{(2)}$.

%----------------------------------------------------------------------------
\subsubsection{Least-square analysis: constraining next-to-leading order (1PN) 
tidal effects from numerical relativity data}
%----------------------------------------------------------------------------

In this subsection we shall firm up the previous analysis and make it
more quantitative by using a $\chi^2$ procedure.
For each EOS, labelled by index $I$, we have $20$ NR data points,
Ref.~\cite{Uryu:2009ye}, $E_b^{\rm Ury{\rm\bar{u}}}(x_{n_I}; \,I)$, where the index
$n_I$ varies from  one to twenty. We retain in our analysis six EOS; 
I=(2H, HB, 2B, FPS, SLy, BGN1H1).
Let us then define the following formal $\chi^2$ function, measuring
the (squared) ``distance'' between NR and EOB: 
\be
\label{eq:chi2}
\chi^2{\left(\bar{\alpha}_1,\delta\right)}=\sum_{I,n}\left[ \dfrac{(E^{\rm Ury{\rm\bar{u}}}_b(x_n; \,I)}{M}
-\delta\, x^4_n) -  \dfrac{E^{\rm EOB}_b(x_n;\,\bar{\alpha}_1,I)}{M}  \right]^2.
\ee
Here, $x=\Omega^{2/3}$ and the index $n$ 
runs (for each EOS label $I$) over the sample of numerical data from one to twenty,
so that $\chi^2$ contains 120 terms in all.
We are interested in studying the dependence of $\chi^2$ over the two
variables $(\delta,\bar{\alpha}_1)$. Here $\delta$ denotes the
coefficient of a 3PN subtraction to NR data of the type that we discussed
in the previous subsection (as motivated by the neglect of some 3PN
terms in the ``waveless'' approximation). 
As explained above, we shall restrict the variation of $\delta$ to the
range $0.4\leq \delta\leq 1.2$. For simplicity, we shall actually sample
this interval through the three  values $\delta=(0.4,0.8,1.2)$.
On the other hand, the coefficient $\bar{\alpha}_1$ parametrizes possible 
next-to-leading order (NLO) 1PN correction to the tidal effects.
We will use the three different descriptions of NLO tidal effects delineated 
in Eqs.~\eqref{eq:linear}-\eqref{eq:harmonic} above.

\begin{figure}[t]
\begin{center}
\includegraphics[width=75 mm, height=58mm]{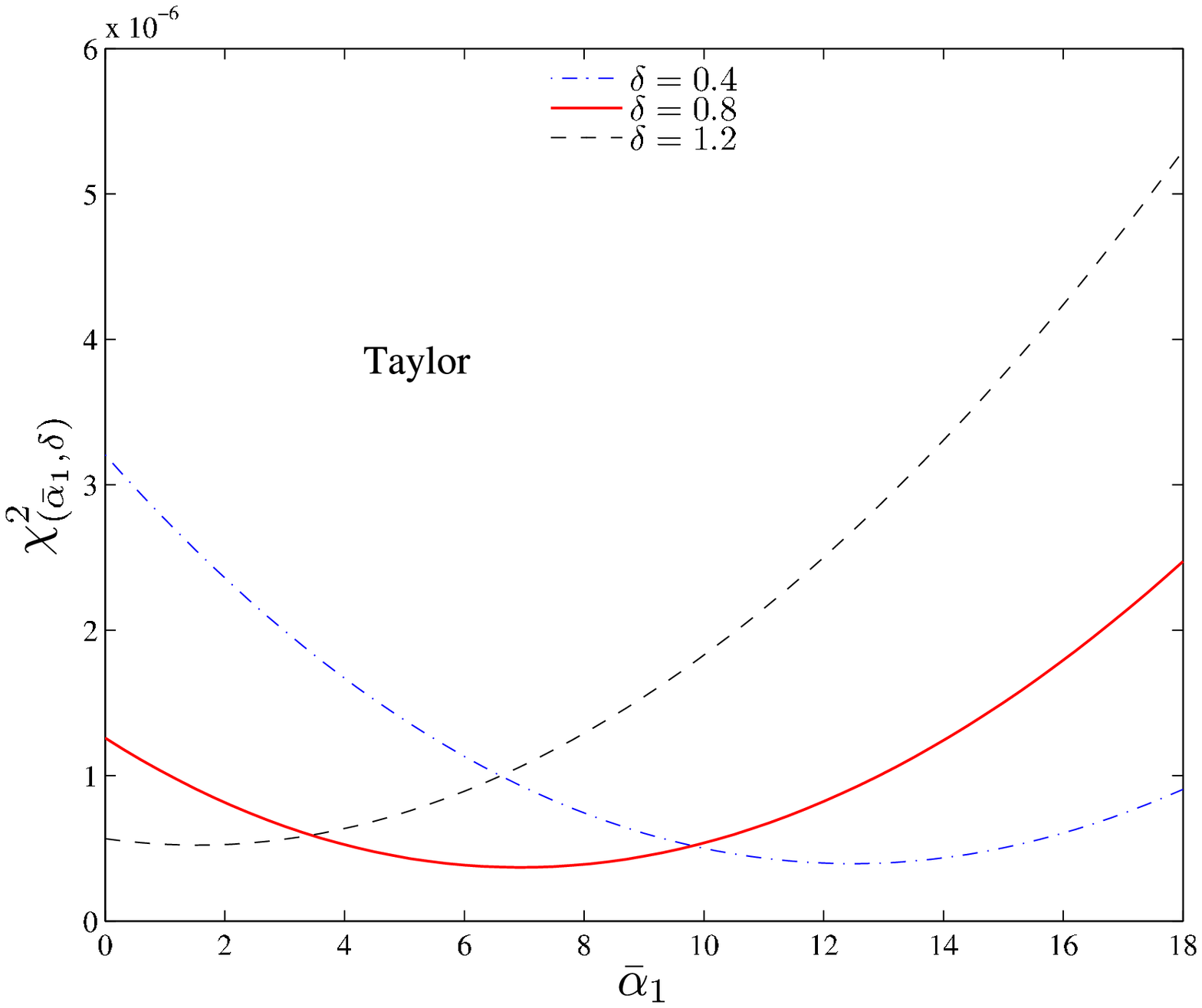}\\
\vspace{2.0mm}
\includegraphics[width=75 mm, height=58mm]{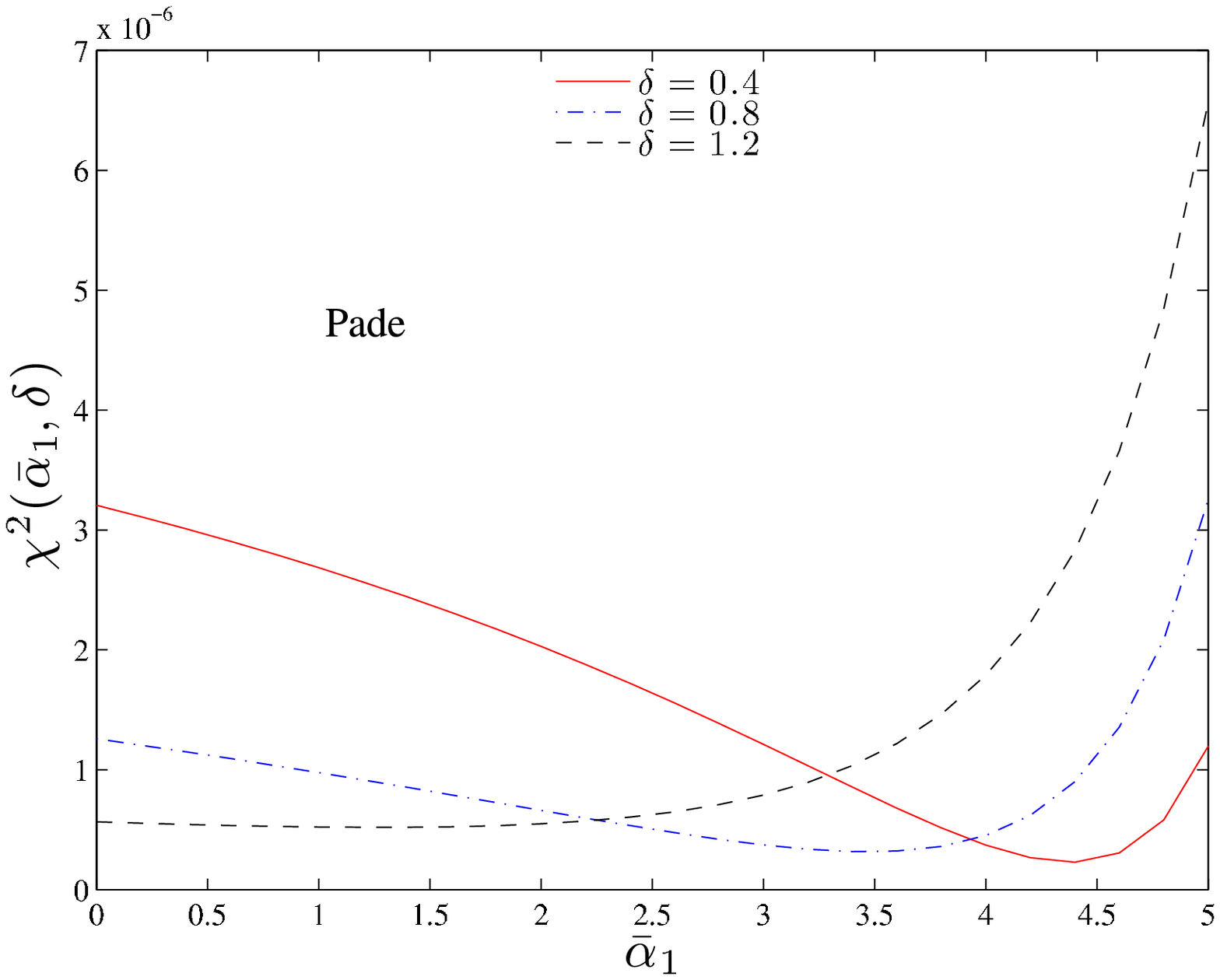}\\
\vspace{2.0mm}
\includegraphics[width=75 mm, height=58mm]{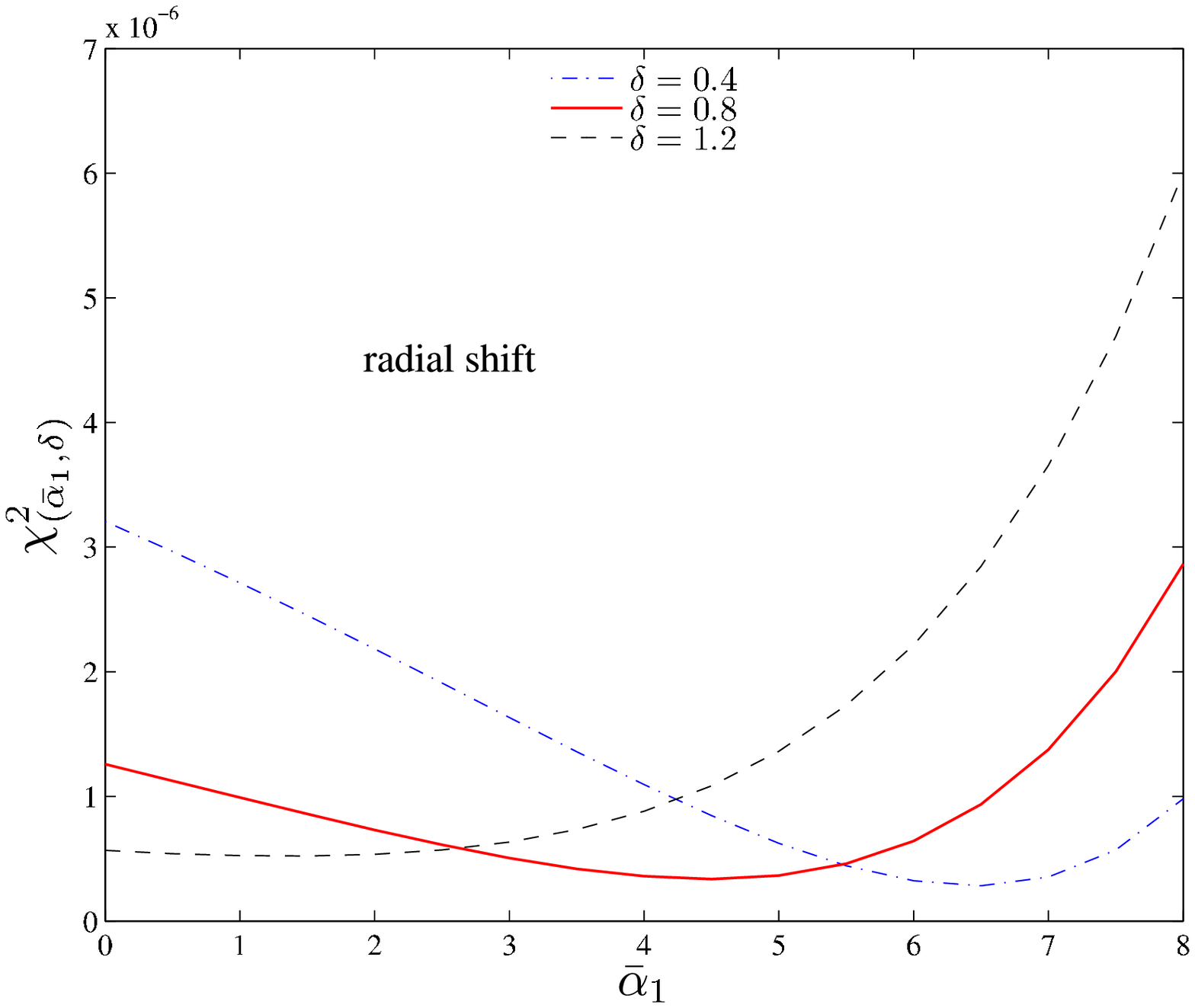}
\caption{\label{fig:fig2} Sections of the function $\chi^2(\bar{\alpha}_1,\delta)$ 
for three values of $\delta$. The figure displays the corresponding ranges of
allowed values of $\bar{\alpha}_1$. Note that, for all models, the minima are
rather shallow.}
  \end{center}
\end{figure}

We wish to use the least-square method, i.e., minimizing the EOB-NR 
``distance'' function $\chi^2{\left(\bar{\alpha}_1,\delta\right)}$,
to constrain the values of $\left(\bar{\alpha}_1,\delta\right)$.
However, we find that $\chi^2{\left(\bar{\alpha}_1,\delta\right)}$ 
remains close (on the scale of the NR error bars) to its global
minimum in a ``valley'' which  extends  over a significant region of 
the $\left(\bar{\alpha}_1,\delta\right)$  plane. 
This means that, given the present error level in numerical data, we
cannot meaningfully and simultaneously select preferred values for 
$\left(\bar{\alpha}_1,\delta\right)$. As a substitute, we shall
exhibit the sections of the $\chi^2$ valley that correspond to the 
three values of $\delta$ selected visually above in Fig.~\ref{fig:fig1}.
In other words, we now fix $\delta$ (to one of its three values) in
Eq.~\eqref{eq:chi2} and consider the dependence of $\chi^2$ on  $\bar{\alpha}_1$.
The resulting one-dimensional plots are exhibited in Fig.~\ref{fig:fig2}.

%-------------------------------
% FIG.2: Chi2 vs alpha_1
%-------------------------------
%--------------------

Each panel of Fig.~\ref{fig:fig2} corresponds to a different modelization
of NLO tidal effects: ``Taylor'' (upper panel,, Eq.~\eqref{eq:linear}), 
``Pad\'e'' (middle panel), Eq.~\eqref{eq:pade} and 
``radial-shift'' (lower panel), Eq.~\eqref{eq:harmonic}. In addition,
each panel contains three curves corresponding to the three 
above-selected values of $\delta$: $\delta=0.4$ 
(dash-dot line, right-most curve), $\delta=0.8$ (solid-line, middle curve),
and $\delta=1.2$ (dashed-line, left-most curve).

Let us start by focussing on the (solid) curves corresponding to 
the ``central'' value of $\delta$, $\delta=0.8$. We see that the
preferred values of $\bar{\alpha}_1$ that they select (minimum of
the curves) are $\bar{\alpha}_1\approx 7$ for the Taylor model, 
$\bar{\alpha}_1\approx 3.5$ for the Pad\'e model and
$\bar{\alpha}_1\approx 4.5$ for the ``radial-shift'' model.
This shows that higher order PN terms (differently included in
the different models) have a significant effect on the determination 
of $\bar{\alpha}_1$.
Note also that when $\delta=1.2$ all the models tend to favour a lower
value: $\bar{\alpha}_1\sim 1$. The value of $\chi^2$ at  $\bar{\alpha}_1=0$
and $\delta=1.2$ is $\chi^2{\left(0,1.2\right)}=5.665\times 10^{-7}$.
This formally corresponds to an average (squared) ``error level'' on the individual
NR-EOB energy differences summed in $\chi^2$ equal to 
$\sqrt{\chi^2{\left(0,1.2\right)}/120}=0.687\times 10^{-4}$. This 
level is comparable to the ``virial error'' on each individual NR
data point $\delta^v E_b/M\sim 10^{-4}$. It is therefore reasonable to
use this level to select a range of values of $\bar{\alpha}_1$.
Combining this range with the range of values of $\delta$'s means
that, at this stage, the range of values of  $\bar{\alpha}_1$ that
is compatible with the NR data is obtained by taking the level 
surface $\chi^2{\left(\bar{\alpha}_1,\delta\right)}=\chi^2{\left(0,1.2\right)}$
as the admissible bottom of the ``valley'' in the 
$\left(\bar{\alpha}_1,\delta\right)$ plane.
This leads to the following admissible ranges: 
$0\lesssim\bar{\alpha}_1\lesssim 15.7$ for the Taylor model;
$0\lesssim\bar{\alpha}_1\lesssim 4.8$ for the Pad\'e model;
$0\lesssim\bar{\alpha}_1\lesssim 7.5$ for the ``radial shift''  model.
It is clear that at this stage the fact that (as we have argued above)
the NR data are ``polluted'' by some systematic errors (notably linked
to unaccounted 3PN effects) prevents us from giving very significant
constraints on the value of $\bar{\alpha}_1$. Note in particular 
that the value $\bar{\alpha}_1=5/4=1.25$ which follows ( in the equal-mass case) 
from Eq.~\eqref{eq:alpha1} is compatible with the present NR data
(if we allow $\delta=1.2$). In this respect, it is interesting to 
note that if we consider a model of the form
\be
\hat{A}^{\rm tidal} = 1 + \bar{\alpha}_1 u + \bar{\alpha}_2 u^2,
\ee
with $\bar{\alpha}_1=1.25$ and compute the corresponding $\chi^2$
for the central value $\delta=0.8$, we find that 
$\chi^2{\left(\bar{\alpha}_2,0.8 \right)}$ reaches a minimum
around $\bar{\alpha}_2\approx 40$. In addition the value of the
minimum of the $\chi^2$ is $3.20\times 10^{-7}$ which is 
slightly better than the performance of the 1PN Taylor model in
the upper panel of Fig.~\ref{fig:fig2}. This shows again that
higher PN tidal effects can play an important role and that
the minimima exhibited (for the central value $\delta=0.8$) 
in the three panels of Fig.~\ref{fig:fig2} should be viewed 
as  ``effective'' values of  $\bar{\alpha}_1$.
We note in this respect that a situation where higher-PN corrections dominate
over the 1PN one is not at all exceptional. For instance, the 1PN contribution
to the EOB radial potential $A(r)$ {\it vanishes}, its 2PN contribution has a
rather small coefficient, $2\nu$, while the numerical coefficient of the 3PN
contribution $\nu a_4$ is quite large and significantly modifies the
conclusions that one might draw from the first two PN contributions. 
%------------------------
% FIG. 3
% Best choices for EOS 2B
%-----------------------
\begin{figure}[t]
\begin{center}
\includegraphics[width=75 mm, height=60mm]{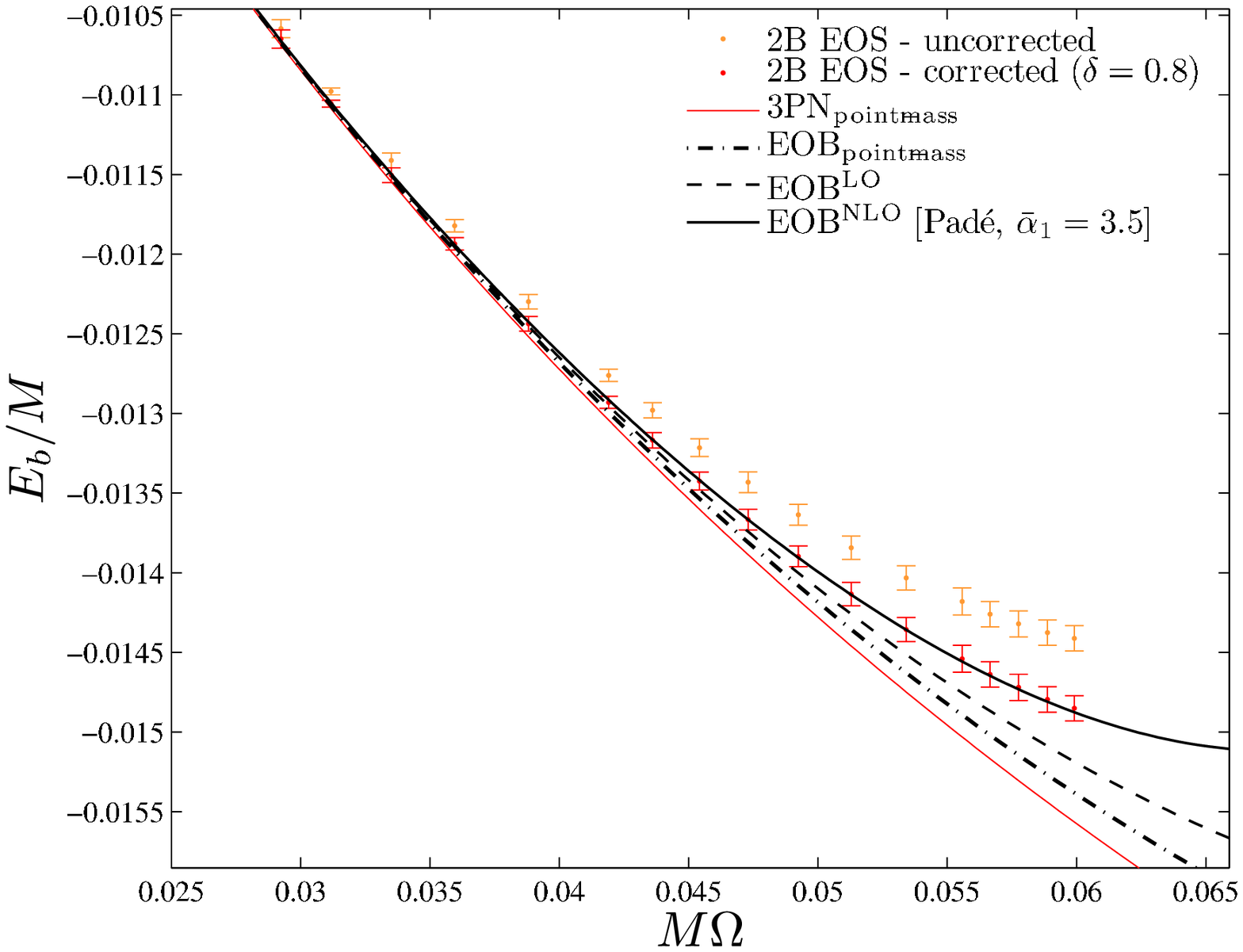}\\
\vspace{2 mm}
\includegraphics[width=75 mm, height=60mm]{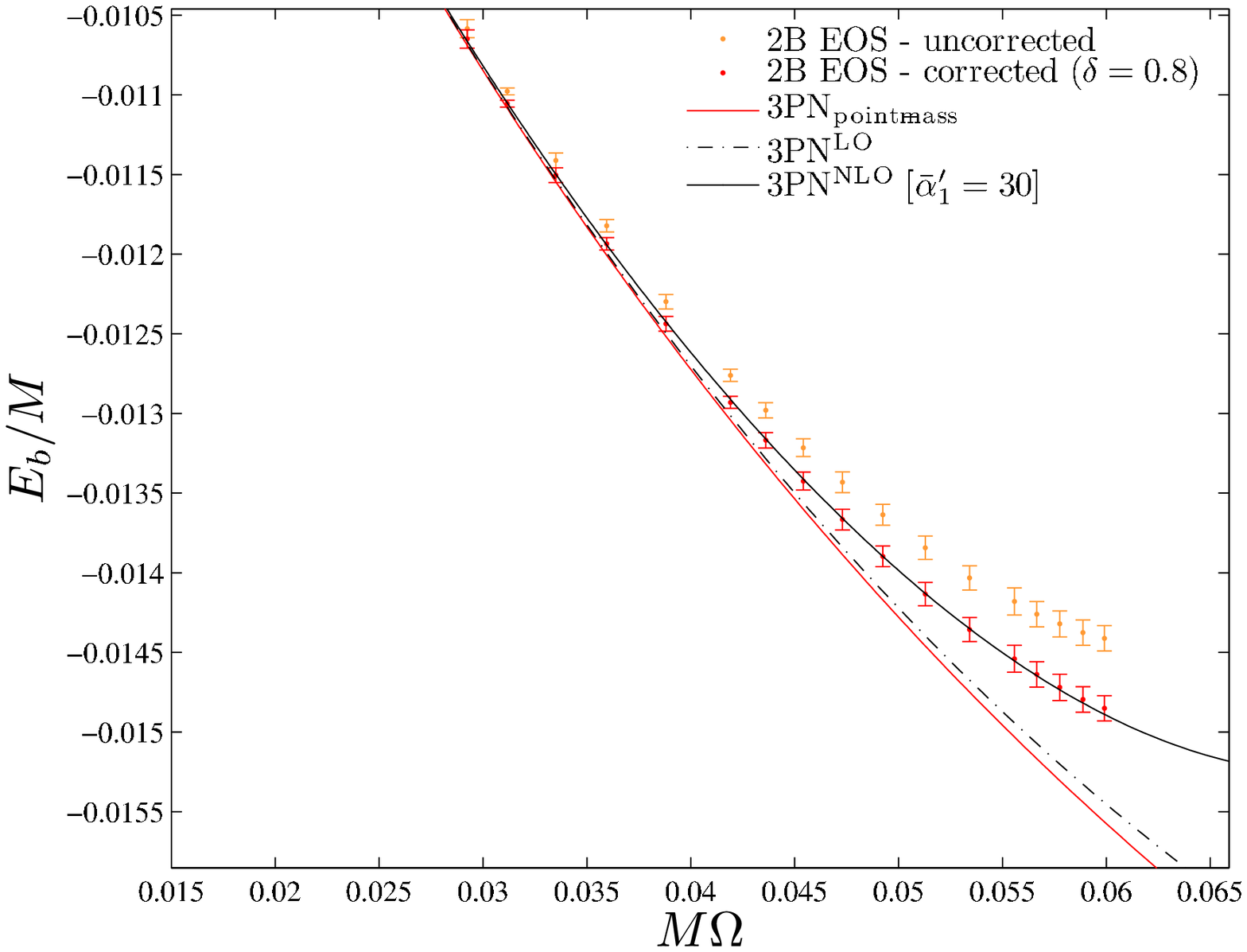}
\caption{\label{fig:fig3}2B EOS: Explicit comparison between various analytical representations
of the binary binding energy and (corrected) numerical relativity data. 
The correction parameter is chosen to be $\delta=0.8$. The upper panel refers to EOB (resummed) 
models. The lower panel to PN (nonresummed) models.  For ${\rm EOB^{NLO}}$  effects, we 
use their Pad\'e representation, Eq.~\eqref{eq:pade} with $\bar{\alpha}_1=3.5$. 
For the 3PN$^{\rm NLO}$ model, we use $\bar{\alpha}_1'=30$. }
  \end{center}
\end{figure}
%--------------------

%------------------------------------------------
% FIG.4 full results for binding energy
%------------------------------------------------
\begin{figure*}[t]
\begin{center}
\includegraphics[width=75 mm, height=60mm]{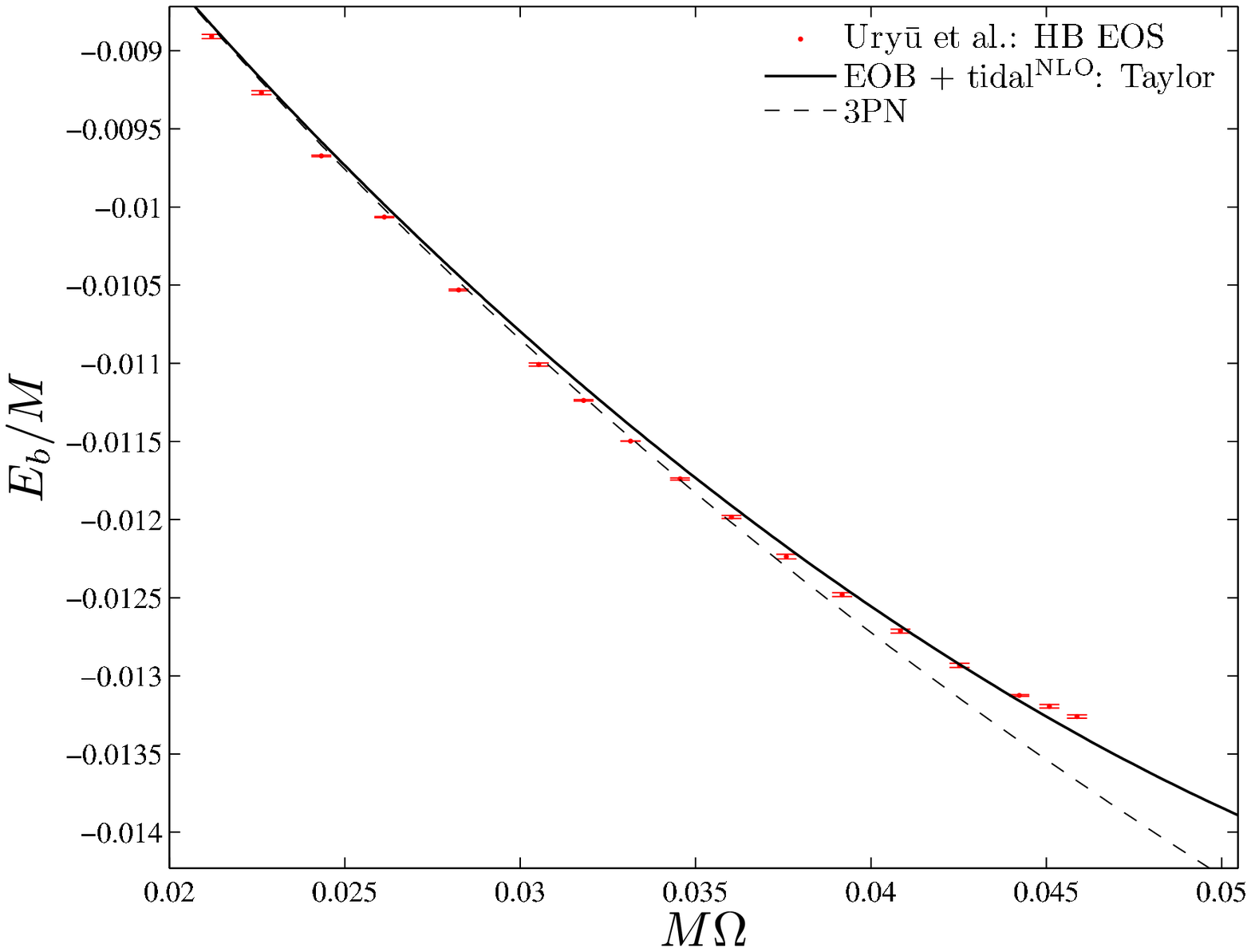}
\hspace{2.5 mm}
\includegraphics[width=75 mm, height=62mm]{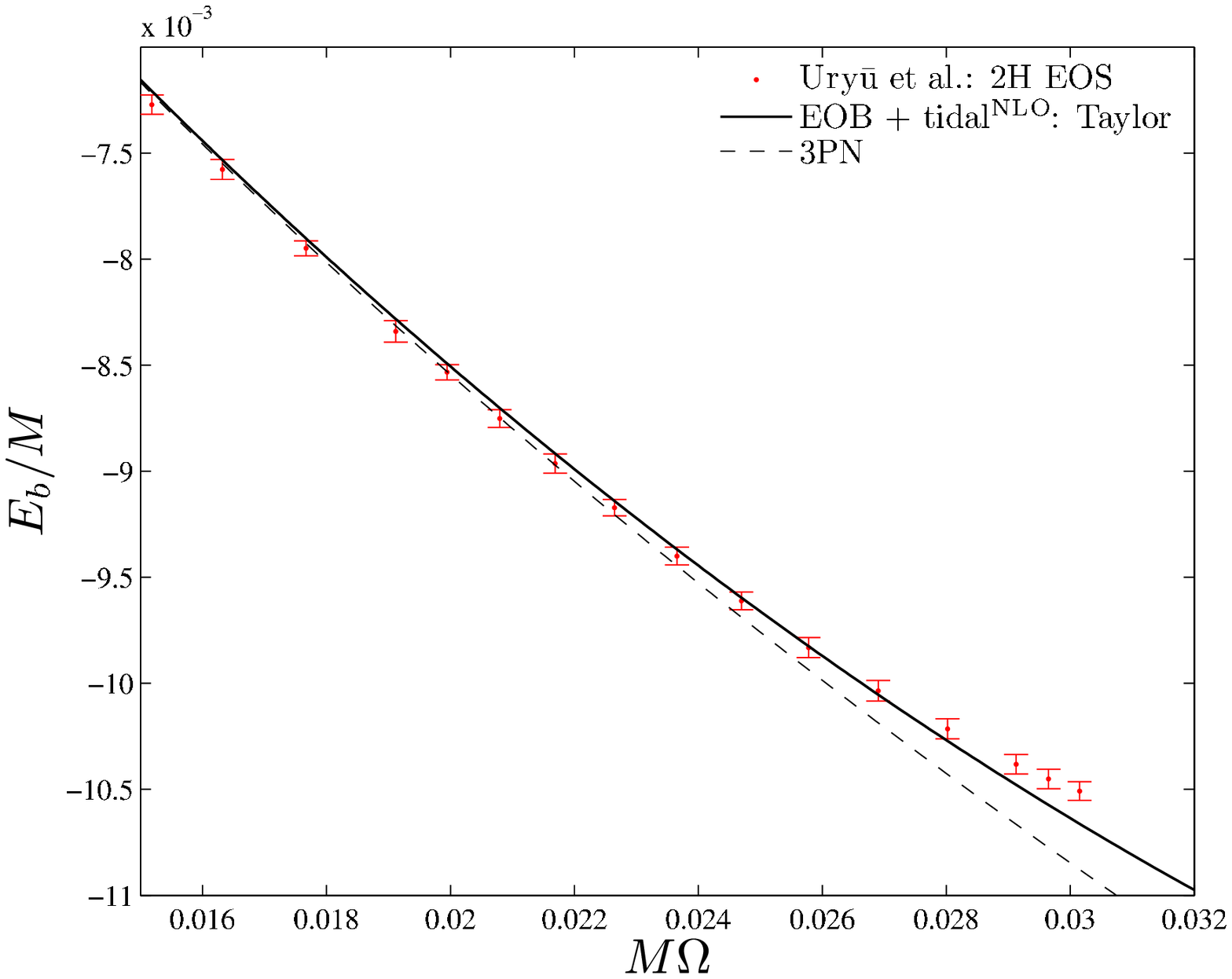}\\
\vspace{3.0 mm}
\includegraphics[width=75 mm, height=60mm]{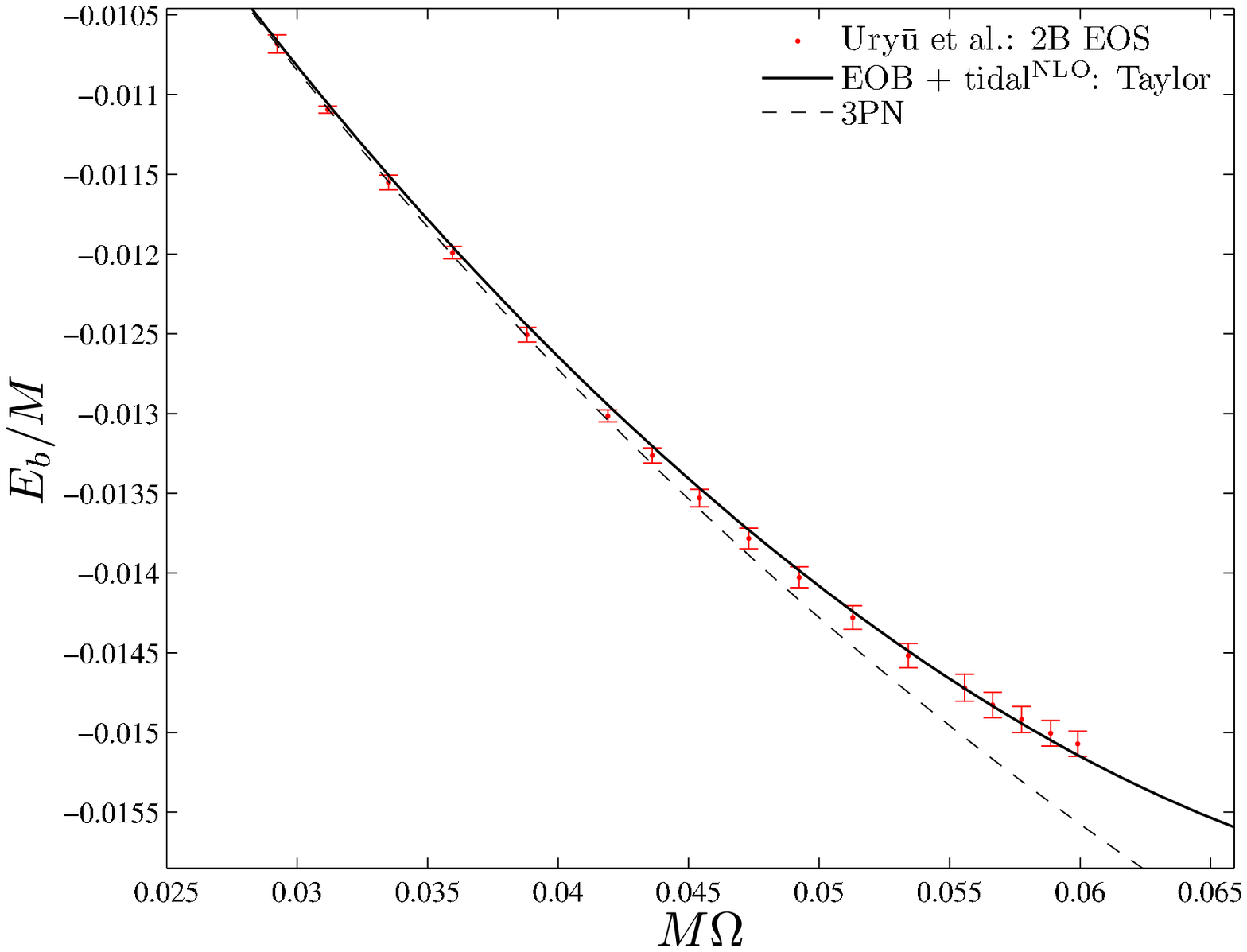}
\hspace{2.5 mm}
\includegraphics[width=75 mm, height=60mm]{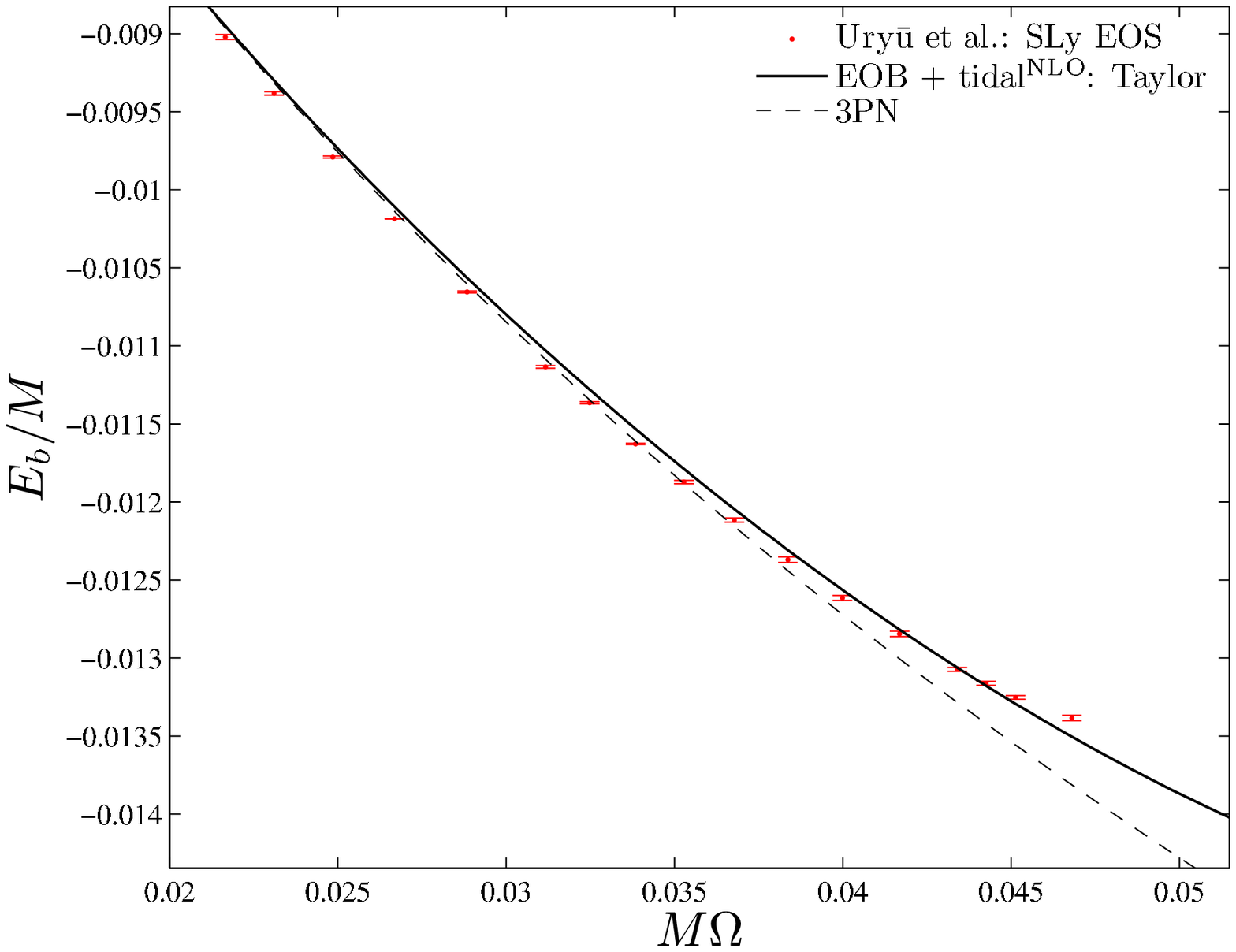}\\
\vspace{3.0 mm}
\includegraphics[width=75 mm, height=60mm]{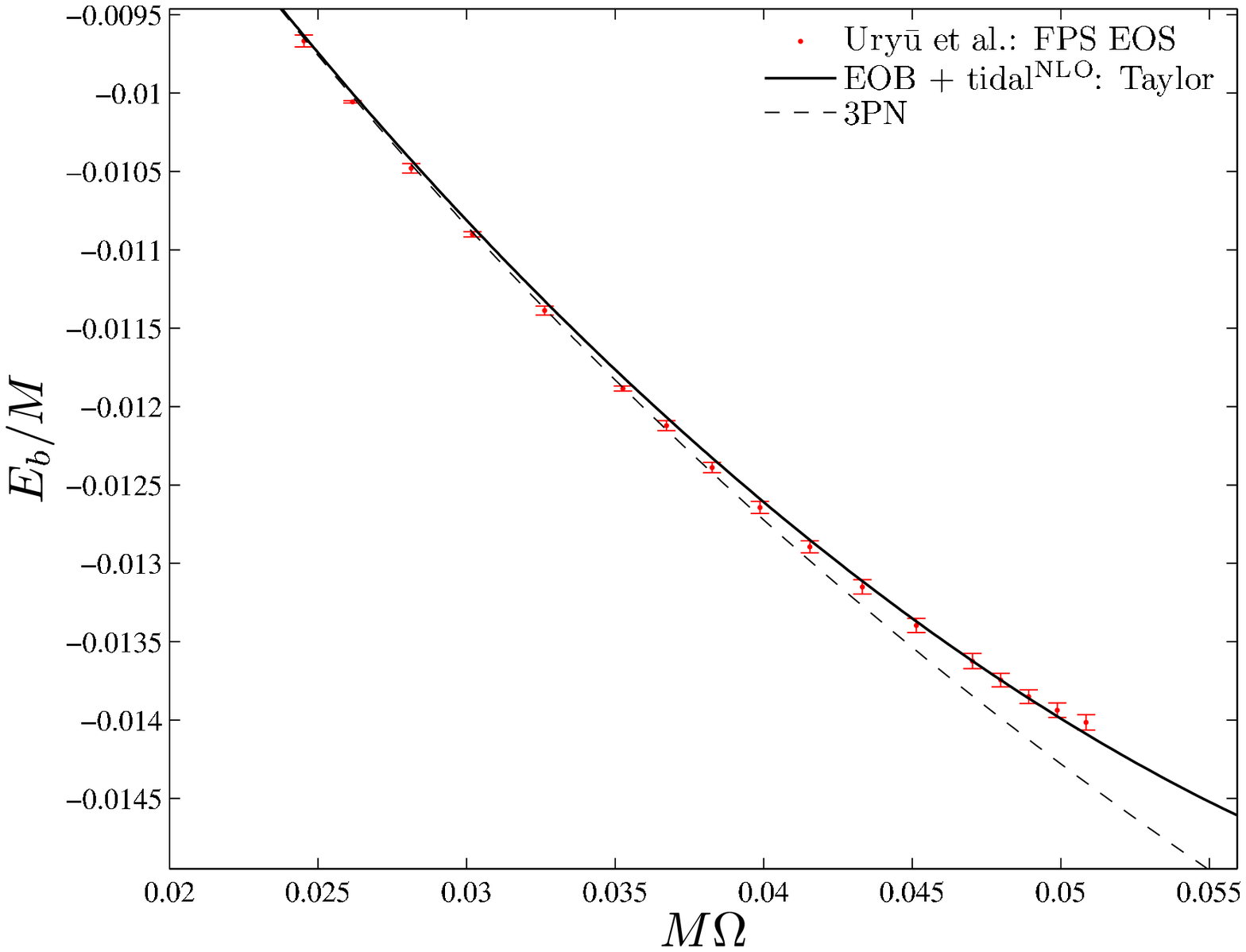}
\hspace{2.5 mm}
\includegraphics[width=75 mm, height=60mm]{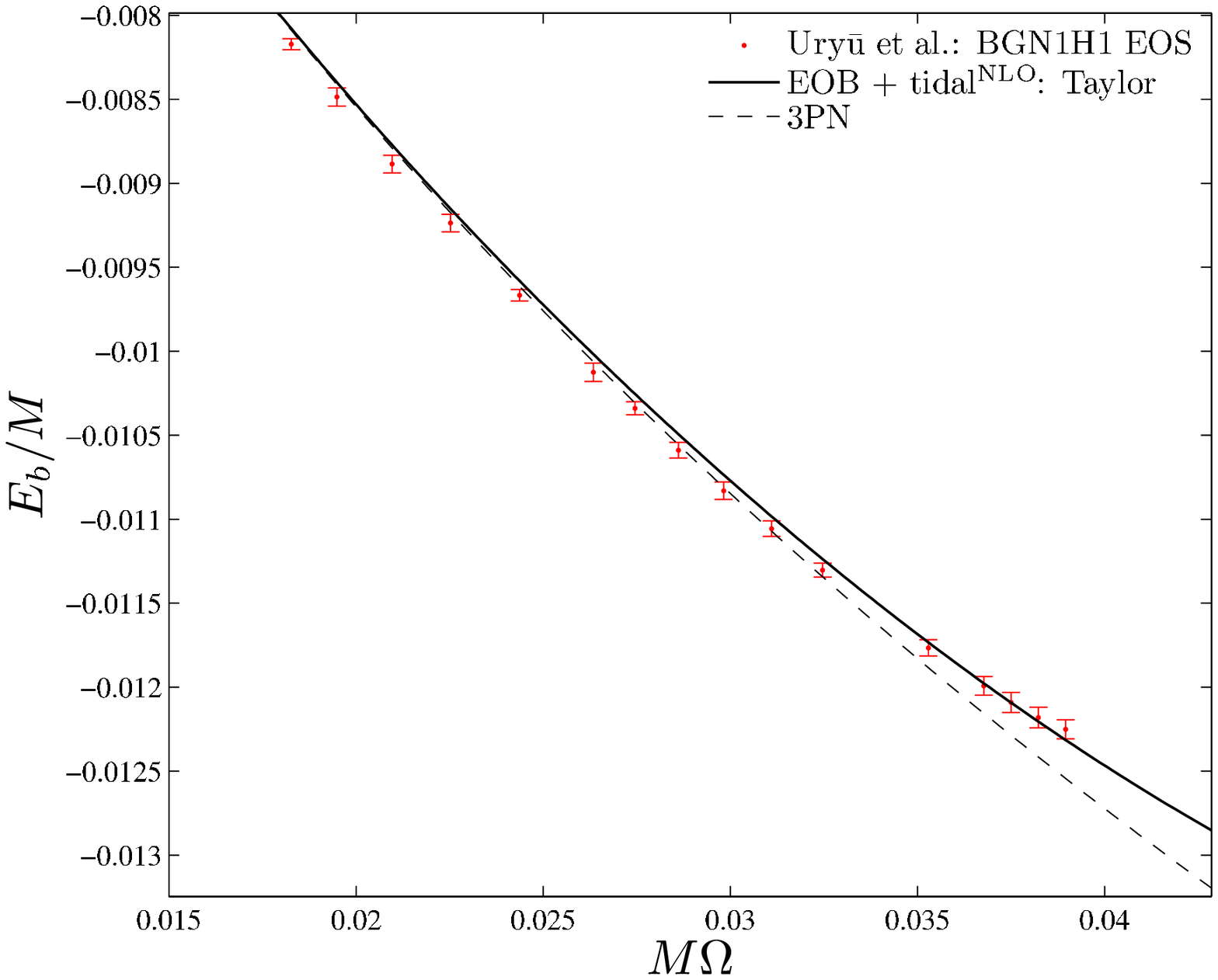}\\
\caption{\label{fig:fig4}Global comparison between EOB$^{\rm NLO}$ and NR binding energies. 
We use the values $(\bar{\alpha}_1,\delta)=(1.25,1.2)$.  The 3PN point-mass curve
is added to guide the eye.}
  \end{center}
\end{figure*}
%--------------------
Figs.~\ref{fig:fig3} and \ref{fig:fig4} illustrate the complementary effects 
of $\bar{\alpha}_1$ and  $\delta$ at the level of the binding energy $E_b$.

Fig.~\ref{fig:fig3} focuses on the 2B EOS model and contrasts (resummed) EOB
(top panel) and (nonresummed) PN analytical representations of the binding energy.
In both cases, the NR binding energy is corrected by the same amount, that is
we assume that $\delta$ takes its central value $\delta=0.8$. 
We see on this figure that the effect of the $\delta$-correction is comparable
to that of the added 1PN tidal contribution.
Note that the value of the $\bar{\alpha}'_1$ parameter needed in the PN 
expanded case (bottom panel) is significantly larger than the ones needed
in the EOB case (for all ways of modeling 1PN tidal contributions).

Fig.~\ref{fig:fig4} illustrates the excellent agreement between the EOB
predictions (here considered for $\bar{\alpha}_1=1.25$)
and the ($\delta$-corrected, with $\delta=1.2$ ) numerical data for
all EOS. The fact that the $\chi^2$ minima exhibited in Fig.~\ref{fig:fig2}
are at a comparable level (for all $\delta$'s in the range we considered) 
indicates that a similarly excellent NR/EOB agreement would have been obtained 
all along an extended valley in the $(\bar{\alpha}_1,\delta)$ plane. In view 
of Fig.~\ref{fig:fig3} the same would hold for the NR/PN agreement, at the
cost, however, of using, on average, significantly larger values of $\bar{\alpha}_1$.

Summarizing: the recent numerical data of Ury${\rm\bar{u}}$ et al. do exhibit
the influence of tidal interactions in close BNS systems. However,
the presence of systematic errors in the data (due to an imperfect satisfaction 
-of the helical-Killing-vector condition) partially masks the tidal interactions
and does not allow for a clean determination of the coefficients
parametrizing tidal effects (and notably their 1PN contributions).

We recommend that new non-conformally flat simulations be performed for
several values of the radius $r_0$ at which the helical-Killing-vector condition
is cut off. By studying the dependence of the results on $r_0$, it might
be possible to extrapolate the results to infinite value of $r_0$
(as used in analytical calculations), and thereby eliminate the 
3PN-level systematic error $ \delta \, x^4$.

%--------------------------------------------------------------------------------------
\section{Incorporating radiative tidal effects in the EOB formalism}
\label{sec:sec4}
%--------------------------------------------------------------------------------------

Besides the specific Hamiltonian (\ref{eq:Heob}), (\ref{eq:Heff}), the other key ingredients of the 
EOB formalism are: (i) a specific, ``factorized'' representation of the multipolar waveforms 
$h_{\l m}$, and (ii) a resummed estimate of the radiation reaction force ${\mathcal F}$, 
which must be added to the conservative Hamiltonian dynamics (\ref{eq:Heob}), (\ref{eq:Heff}). 
In the most recent, and seemingly most accurate, version of the EOB formalism the radiation reaction 
is analytically computed in terms of the multipolar waveforms. Therefore, it will be enough to 
estimate here the ``tidal correction'' to the multipolar waveforms $h_{\ell m}$. Following the 
``factorization'' philosophy of Refs.~\cite{Damour:2009vw,Damour:2008gu} we shall look 
for tidal-correction factors  $f_{\l m}^{\rm tidal} = 1 + {\mathcal O} (\mu , \sigma)$, such that 
the EOB waveform would read
\be
\label{eq:4.1}
h_{\l m} = f_{\l m}^{\rm tidal} \, h_{\l m}^0.
\ee
Here $h_{\l m}^0$ is the factorized BBH EOB waveform, 
introduced in~\cite{Damour:2008gu}, and augmented by
two next-to-quasi-circular parameters $(a_1,a_2)$ in
Ref.~\cite{Damour:2009kr}. 
[Note, however, that, in view of the smallness of the tidal effects on the waveform,
 $f_{\l m}^{\rm tidal} - 1 \ll 1$, it would be equivalent to use (as done for the $A(r)$
potential) an {\it additive} ansatz: $h_{\l m} = h_{\l m}^0 + h_{\l m}^{\rm tidal}$.]

In principle, one can use the effective action (\ref{eq:2.3}), (\ref{eq:2.4}) to compute the tidal 
contributions to the waveform with any required relativistic accuracy (post-Minkowskian and/or 
post-Newtonian).

Here, we shall focus on the leading PN-order tidal correction to the leading PN waveform, i.e. 
the $\l = 2$, $m=2$ partial wave $h_{22}$. This will provide the leading tidal correction to the 
radiation reaction (which is predominantly given by a contribution $\propto \vert 2 \, \Omega \, h_{22} 
\vert^2$).

In that case, a shortcut for computing the tidal correction $f_{22}^{\rm tidal}$ consists in noting that the 
quadrupolar gravito-electric contribution in the action (\ref{eq:2.3}) corresponds to adding to the 
energy-momentum tensor of point-masses an extra contribution $\Delta \, T_{(x)}^{\mu\nu} \equiv 
2 \, g^{-1/2} \, \delta \Delta_{\rm nonminimal} \, S / \delta g_{\mu\nu}$, which describes the tidally 
induced quadrupole moment in each body $A$. At the leading ``Newtonian'' order this means that 
the quadrupole mass moment $M_{ij}$ of the system will be
\be
\label{eq:4.2}
M_{ij} = \sum_A {\rm STF}_{ij} [M^A  z_A^i \, z_A^j + \mu_2^A \, G_{ij}^A] \, ,
\ee
where STF denotes a symmetric trace-free projection, and where the second term is the tidally 
induced quadrupole moment. Replacing the Newtonian value (\ref{eq:2.6}) of $G_{ij}^A$ 
(computed using Eq.~(\ref{eq:2.7})) yields
\begin{eqnarray}
\label{eq:4.3}
M_{ij} &= &\sum_A {\rm STF}_{ij} \left[ M_A \, z_A^i \, z_A^j + 3\mu_2^A \, GM_B \, \frac{z_{AB}^i \, 
z_{AB}^j}{r_{AB}^5} \right] \nonumber \\
&= &\left( \mu + \sum_A 3 \, \mu_2^A \, \frac{GM_B}{r_{AB}^5} \right) r_{AB}^2 \, \hat n_{AB}^{ij} \, ,
\end{eqnarray}
where $\mu \equiv M_A \, M_B / (M_A + M_B)$ is the reduced mass of the binary system, and where 
we reduced the first expression to the center-of-mass
frame. Eq.~\eqref{eq:4.3}  agrees with 
Eq.~(7) of~\cite{Flanagan:2007ix} (in the limit where one neglects the
excitation of the internal radial modes: $x_n \to 0$). In 
addition to the explicit tidal modification $\propto \mu_2^A$ that appears in the first factor of 
Eq.~(\ref{eq:4.3}), there is an implicit tidal effect coming from the fact that the EOB waveform is 
conventionally expressed in terms of the (instantaneous) orbital frequency $\Omega$ of the binary 
system. We must then eliminate the relative distance $r_{AB}$ in Eq.~\eqref{eq:4.3} in favor of 
$\Omega$. This is done by using the adiabatic (quasi-circular) Kepler law. The latter is modified by 
tidal forces:
\begin{eqnarray}
\label{eq:4.4}
\Omega^2 z_{AB}^i &= &- \frac{d^2 z_{AB}^i}{dt^2} = -\frac{1}{\mu} \, \frac{\partial L}{\partial z_{AB}^i} 
\nonumber \\
&= &\frac{GM}{r_{AB}^3} \, z_{AB}^i - \frac{1}{\mu} \, \frac{\partial L^{\rm tidal}}{\partial z_{AB}^i} \, .
\end{eqnarray}

Differentiating the leading $(\ell = 2)$ tidal Lagrangian \eqref{eq:2.14}, and keeping only the leading 
$(\ell = 2)$ term yields a modified Kepler law of the form
\be
\label{eq:4.5}
\Omega^2 r_{AB}^3 = GM \left[ 1+9 \, \frac{M_B}{M_A} \, \frac{G\mu_2^A}{r_{AB}^5} + 9 \, \frac{M_A}
{M_B} \, \frac{G\mu_2^B}{r_{AB}^5} \right] .
\ee
Using (\ref{eq:4.5}) to solve $r_{AB}$ in terms of $\Omega$, 
and replacing the (tidally-corrected) answer in \eqref{eq:4.3} 
finally leads to a quadrupole moment of the form
\be
\label{eq:4.6}
M_{ij} = f_{22}^{\rm tidal} \mu \, r_{AB}^2 \, \hat n_{AB}^{ij}
\ee
with a tidal-correction factor
\begin{eqnarray}
\label{eq:4.7}
f_{22}^{\rm tidal} &= &1 + \sum_A \, 3 \, \frac{G\mu_2^A}{r_{AB}^5} \left( \frac{M_B}{\mu} + 
2 \, \frac{M_B}{M_A} \right) \nonumber \\
&= &1 + \sum_A \, 3 \, \frac{G\mu_2^A}{r_{AB}^5} \left( 1 + 3 \, \frac{M_B}{M_A} \right) \nonumber \\
&= &1 + \sum_A \, 2 \, k_2^A \left( \frac{R_A}{r_{AB}} \right)^5 \left( 1 + 3 \, \frac{M_B}{M_A} \right) \, .
\nonumber \\
\end{eqnarray}

The factor $f_{22}^{\rm tidal}$ is the $\l = 2$, $m=2$ tidal-correction factor which was introduced in 
Eq.~(\ref{eq:4.1}). It remains, however, to eliminate $r_{AB}$ in terms of $\Omega$, or, as used in 
the waveform of Ref.~\cite{Damour:2008gu}, in terms of the EOB variable 
$v_{\Omega} \equiv r_{\Omega} \, \Omega$  introduced in~\cite{Damour:2006tr}: 
at the leading order it is enough to use 
$GM / c_0^2 \, r_{AB} = v_{\Omega}^2 (1+{\mathcal O} (1/c_0^2))$. This yields
\begin{align}
\label{eq:4.8}
&f_{22}^{\rm tidal} = 1 \nonumber\\
& + \left( \sum_A \, 2 \, k_2^A \left( \frac{R_A \, c_0^2}{G(M_A + M_B)} \right)^5 
\left( 1+3 \, \frac{M_B}{M_A} \right)\right) v_{\Omega}^{10} \, .  
\end{align}
The result (\ref{eq:4.8}) agrees (after squaring it) with Eq.~(8c) of 
Ref.~\cite{Flanagan:2007ix} (in the limit $x_n \to 0$).

Summarizing: we propose to incorporate radiative tidal effects in the EOB formalism by 
inserting  in the dominant $\l = 2$, $m=2$
waveform, a factor of the form 
\begin{align}
\label{eq:4.8pn}
f_{22}^{\rm tidal} &=  1 \nonumber\\
&+ \left( \sum_A \, 2 \, k_2^A \left( \frac{R_A \, c_0^2}{G(M_A + M_B)} \right)^5 
\left( 1+3 \, \frac{M_B}{M_A} \right)\right) v_{\Omega}^{10}\nonumber\\
&\times\left(1+\beta_1 v_{\Omega}^2\right) \, ,
\end{align}
where we included a possible 1PN correction to radiative tidal effects.
One then computes a tidal-corrected radiation reaction by using 
this corrected waveform in the definition of ${\mathcal F}$ 
given in~\cite{Damour:2008gu} and~\cite{Damour:2009vw}. In principle the
(mass-ratio dependent) coefficient $\beta_1$ can computed analytically. It can
also be ``calibrated'' by comparing NR data of inspiralling BNS systems to
the EOB predictions.

%--------------------------------------------------------------------------------------
\section{EOB predictions for the motion and radiation of inspiralling compact binaries}
\label{sec:sec5}
%--------------------------------------------------------------------------------------

Having defined a specific EOB way of incorporating tidal effects in the motion and radiation of 
inspiralling compact binaries (BNS or BHNS) let us study the predictions made by the resulting 
tidally-extended EOB formalism.

\subsection{Adiabatic inspiral, ``last stable orbit'', and ``contact''}

Let us start by considering the {\it adiabatic} approximation to the inspiral, i.e. the approximation in 
which the inspiral is described as a sequence of circular orbits. In this approximation, a key concept 
is that of the Last Stable (circular) Orbit (LSO).
We saw above the equation determining, in the EOB formalism, the 
sequence of circular orbits, Eq.~\eqref{eq:5.2}.
For large values of $p_{\varphi}$, and large values of $r$ (i.e. small values of $u = 1/r$), 
Eq.~(\ref{eq:5.2}) has a unique solution $r = 1/u \simeq p_{\varphi}^2$, corresponding to Newtonian 
circular orbits. However, when $p_{\varphi}^2$ decreases (as it does along the sequence of 
inspiralling orbits driven by radiation reaction), the sequence of stable circular orbits will 
terminate at certain values $r_{\rm LSO} \equiv 1/u_{\rm LSO}$, $p_{\varphi_{\rm LSO}}^2$ where 
there exists a double root of Eq.~(\ref{eq:5.2}), i.e. a common root of Eq.~(\ref{eq:5.2}) and
\be
\label{eq:5.3}
A''(u) + p_{\varphi}^2 \, B''(u) = 0 \, .
\ee
The condition determining the radial location of the Last Stable Orbit (LSO)
is the vanishing of the determinant
\begin{eqnarray}
\label{eq:5.4}
\left\vert \begin{matrix} A' &B' \\ A'' &B'' \end{matrix} \right\vert_{\rm LSO} &= &A' (u_{\rm LSO}) \, 
B'' (u_{\rm LSO}) \nonumber \\
&- &A'' (u_{\rm LSO}) \, B' (u_{\rm LSO}) = 0 .
\end{eqnarray}
For instance, in the test-mass limit, and in absence of tidal corrections, i.e. for $A(u) = 1-2u$, 
$B(u) = u^2 \, A(u) = u^2 - 2 \, u^3$, Eq.~(\ref{eq:5.4}) reads $-4 \, (1-6 \, u_{\rm LSO}) = 0$, so that 
we recover the classic result $r_{\rm LSO} = 1/u_{\rm LSO} = 6$ (i.e. $r_{\rm LSO}^{\rm phys} = 
6 \, GM$) for the LSO around a Schwarzschild black hole. On the other hand, when inserting in 
Eq.~(\ref{eq:5.4}) 
the complete value of the $A$ function, i.e. the sum (\ref{eq:3.5}), where $A^0 (r;\nu)$ is given 
by Eq.~(\ref{eq:3.4}), and $A^{\rm tidal} (r)$ by Eq.~(\ref{eq:3.6}), we see that the LSO predicted by 
the EOB formalism will depend both on the symmetric mass ratio $\nu$, and on the EOB tidal 
constants $\kappa_\l^{\rm T}$, Eq.~(\ref{eq:3.7}). More precisely, these two types of effects (the 
$\nu$-dependent ones which exist already in BBH systems, and the tidal-dependent ones which exist 
only in BHNS and BNS systems) act in opposite directions. Indeed, the $\nu$-dependent 
contributions tend to make the radial potential $A(r)$ less attractive (see Eq.~(\ref{eq:3.3})), while 
the tidal ones make $A(r)$ more attractive. As a consequence, $\nu$-effects tend to move the radial 
location of the LSO towards smaller values ($r_{\rm LSO} (\nu) < 6 \, GM$), while tidal effects tend to 
move $r_{\rm LSO}$ towards larger values. To avoid gauge effects, it is convenient to measure the 
location of the (adiabatic) LSO in terms of the corresponding (real) orbital frequency
\be
\label{eq:5.5}
\Omega = \frac{\partial H_{\rm EOB}}{\partial \, p_{\varphi}^{\rm phys}} = \frac{1}{GM\mu} \, \frac{\partial 
H_{\rm EOB}}{\partial \, p_{\varphi}} \, .
\ee
Finally, we conclude that the dimensionless orbital frequency $GM\Omega$ at the LSO is a 
function of the dimensionless parameters $\nu$, $\kappa_\l^{\rm T}$ which tends to 
{\it increase} as $\nu$ increases, and to {\it decrease} as $\kappa_\l^{\rm T}$ increases. 
We have seen above that the tidal coefficients $\kappa_\l^{\rm T}$ generically take rather large 
numerical values, of order $\kappa_2^{\rm T} = {\mathcal O} (100)$, when
$\ell=2$,  see Table~\ref{tab:table1}. 
However, they enter the $A$ function at a higher order in $u$ than the
$\nu$-dependent effects. As a consequence, 
the combination of the influences of $\nu$ and $\kappa_2^{\rm T}, \kappa_3^{\rm T} , \ldots$ leads to orbital LSO 
frequencies which are sometimes larger, and sometimes smaller than the ``Schwarzschild value'' 
$GM\Omega_{\rm Schw} = 6^{-3/2} = 0.06804$. This is illustrated in
Table~\ref{tab:table3}  
which lists the values of {\it twice} the orbital frequency 
(corresponding to the adiabatic gravitational wave 
frequency $\omega_{\l m}$ for the dominant mode $\l = m = 2$) for several compactnesses 
($0.13$, $0.17$, $0.17385$, $0.5$) and for two paradigmatic systems: an equal-mass BNS system 
and a binary black hole system (labelled by its formal compactness $c=0.5$).
Here we took the piece-wise polytropic SLy  EOS. Note that one NS mass is
smaller than the ``canonical'' $1.35M_\odot$ so to explore a smaller compactness.
If needed, one can convert the dimensionless freqency  $2 \, GM \Omega$ in Hz
by using $GM_{\odot} = 4.925490947 \mu $ s ($= 1.476625038$~km )  so that 
the conversion factor between $\hat\omega = GM \omega$ and $f = \omega / 2\pi$ is
\be
\label{eq:5.6}
f = \frac{\hat \omega}{2\pi \, GM} = 32.3125 \, \hat\omega \left( \frac{M_{\odot}}{M} \right) {\rm kHz} \, .
\ee
We see that, in a BNS system, the LSO frequency is smaller than the
``Schwarzschild value'' $2 \, GM \Omega_{\rm Schw} =1/(3\sqrt{6})= 0.136083$
for compactness smaller than about 0.1704. For such system the radius of 
the LSO is larger than the canonical Schwarzschild 6GM.
Note, by comparing BNS to BBH ones, how tidal effects can significantly change
the LSO frequency by more than a factor two! The results shown in Table~\ref{tab:table3}
have been computed using the leading order, non-PN-corrected EOB description
of tidal effects. The inclusion of next-to-leading order effects, notably with
$\bar{\alpha}_{1}\sim 6$, would double the effect of tidal interactions at the
LSO and would therefore significantly affect the numbers listed in the table.

%==========
% Table 3
%==========
\begin{table*}[t]
\caption{\label{tab:table3} Adiabatic LSO information for BNS and BBH systems.
The NS models are built using the piece-wise polytropic SLy EOS. From left to 
rigth, the columns report: the composition of the binary, the compactness $c$
of the objects, the NS mass $M$, the NS radius $R$, twice the orbital
frequency at the adiabatic (EOB) LSO $2GM\Omega_{\rm LSO}^{\rm adiab}$,
the corresponding LSO radius $r^{\rm contact}/GM$, the ``contact'' frequency 
$2GM\Omega^{\rm contact}$ and the corresponding radial distance $r^{\rm contact}/GM$.  }
\begin{center}
  \begin{ruledtabular}
  \begin{tabular}{lcccccccc}
    System   &    $c$   & $M$ [$M_\odot$] & $R$ [km] &  $2GM\Omega_{\rm
   LSO}^{\rm adiab}$ & $r_{\rm LSO}/GM$ & $2GM\Omega^{\rm contact}$
   & $r^{\rm contact}/GM$   \\
   \hline \hline
 BNS$^{\rm LO}$                & $0.13$      & 1.0050   &  11.417  & 0.10208  & 7.3991 &0.09590  & 7.6923 &  \\
 BNS$^{\rm LO}$                           & $0.17$         & 1.3205   &  11.470  & 0.13605  & 6.0111 & 0.14060  & 5.8824  \\
 BNS$^{\rm LO}$                           & $0.17385$   & 1.35     &  11.466 & 0.13902 & 5.9163 & 0.145061  & 5.7521  \\
\hline
 BNS$_{\bar{\alpha}_1=7.0}^{\rm NLO}$        & $0.13$      & 1.0050   &  11.417  & 0.09056  & 8.1698 & 0.09834  & 7.6923  \\
 BNS$_{\bar{\alpha}_1=7.0}^{\rm NLO}$   & $0.17385$   & 1.35     &  11.466  & 0.12185  & 6.5120 & 0.148750  & 5.7521  \\
\hline
 BBH$_{\nu=1/4}$    & $0.5$       & $\dots$  &  $\dots$ & 0.19285  & 4.6186 & $\dots$ & $\dots$   \\
 BBH$_{\nu=0}$       & $0.5$       & $\dots$  &  $\dots$ & 0.13608  & 6.0000 & $\dots$ & $\dots$    
  \end{tabular}
\end{ruledtabular}
\end{center}
\end{table*}%

In some BNS systems the concept of LSO and LSO frequency has only a formal meaning 
because the two NS's enter in contact (slightly) before reaching the LSO. This is
illustrated in Table~\ref{tab:table3} 
 which lists also the value of (twice) the orbital frequency at the moment of ``contact'', i.e. when 
the EOB radial separation $R$ becomes equal to the sum of the two (areal) radii $R_A + R_B$. [We 
use $R_B = 2 \, GM_B$ when the companion is a BH.] 
Note that it is approximately given by the simple analytical formula
\be
2GM\Omega_{\rm contact}\approx 2\left(\dfrac{X_A}{c_A} + \dfrac{X_B}{c_B}\right)^{-3/2}.
\ee
This definition of ``contact'' relies on the use of 
the EOB radial coordinate. As this coordinate is a smooth deformation of the usual areal coordinate, 
we think that it is a reasonable definition, and we propose here to use the EOB description up to the 
moment when either the two objects enter in contact, or (if it happens earlier) when the orbital 
frequency $\Omega$ reaches a maximum. Note also that Table~\ref{tab:table3}
illustrates the possible effect  (for $c=0.13$) of NLO (1PN) tidal
contributions. 
This effect is very significant. The second line of the table indicates that, when using a
``Taylor'' model with $\bar{\alpha}_1=7.0$ (which was the minimum of $\chi^2$
for the central value of $\delta$), the arrangement of the LSO and touching
radius changes. In absence of 1PN correction the contact was reached {\it before} LSO,  
while with $\bar{\alpha}_1=7.0$ the contact is reached {\it after}
the LSO, which means that the system undergoes a short ``plunge phase'' 
before entering in contact.

In addition to the discussion of the frequency at the moment of contact
(i.e. when $R = R_A + R_B$) let us also consider the 
dimensionless parameter measuring the 
tidal deformation of the NS labelled $A$ by its companion $B$
\be
\label{eq:5.7}
\epsilon_A = \frac{M_B}{R^3} \, \frac{R_A^3}{M_A} \, .
\ee
At contact, $(R = R_A + R_B)$, this parameter can be expressed in terms of the two compactnesses 
$c_A = GM_A / R_A$ and $c_B = GM_B / R_B$ as
\be
\label{eq:5.8}
\epsilon_A^{\rm contact} = \frac{c_B}{c_A} \, \frac{R_A^2 \, R_B}{(R_A + R_B)^3} \, .
\ee
For a symmetric, equal-mass BNS system, we see that, upon contact, $\epsilon_A^{\rm contact} = 
\epsilon_B^{\rm contact} = 1/8$. It was found in~\cite{Damour:2009vw}, and briefly recalled above, 
that the fractional deformation of the NS $A$ is given by the product $h_2^A \, \epsilon_A$, where 
the ``shape'' Love number $h_2^A$ is of order $0.8$ for a typical NS compactness. 
This means that, in a symmetric (or near symmetric) BNS system each NS is only deformed by 
about $10\%$ at the moment of contact. This motivates our proposal of using the EOB description 
up to the moment of contact.

In the case of asymmetric BHNS systems (with $A$ labelling the NS and $B$ the BH) we can reach a 
similar general conclusion by noticing that the dimensionless function $R_A^2 \, R_B / (R_A + 
R_B)^3$ (which depends only on the ratio $R_A / R_B$) reaches a maximum value of $2^2 / 3^3 = 
4/27$ when $R_A = 2 \, R_B$. As a consequence, we have the general inequality
\be
\label{eq:5.9}
\epsilon_A^{\rm contact} \leq \frac{4}{27} \, \frac{c_B}{c_A} \, .
\ee

In the present case, $B$ denotes a BH (with $c_B = \frac{1}{2}$) so that $\epsilon_A^{\rm contact} \leq 
2/(27 \, c_A) = 0.074074 / c_A$. Upon multiplication by $h^A_2 \sim 0.8$ this yields 
$h^A_2 \, \epsilon_A^{\rm contact} \lesssim 0.06 / c_A$. As NS compactnesses are expected to be 
larger than about $0.13$, we find that the NS in a BHNS system is expected to be always deformed 
by less than $50\%$ up to the moment of ``contact'' with its BH companion. Actually, the reasoning 
above shows that such large deformations are only attained when $R_A = 2 \, R_B$, i.e. 
when the mass ratio is equal to
\be
\label{eq:5.10}
\frac{M_B}{M_A} = \frac{c_B}{c_A} \, \frac{R_B}{R_A} = \frac{1}{2} \, \frac{c_B}{c_A} = \frac{1}{4 \, c_A} \, .
\ee
For typical NS compactnesses $c_A \sim 0.15$, such a mass ratio $M_B / M_A \sim 1.67$ would 
correspond to a BH of a small mass ($M_B \sim 2.3 \, M_{\odot}$). Larger BH masses will lead to 
smaller deformations of the NS.

Summarizing, the main conclusions of this subsection are that: (i) the EOB
formalism predicts that the ``quasi point mass'' description can be applied 
up to contact, without the possibility of a 
disruption of the NS's in a well detached state, and (ii) the divide between
the systems that undergo a plunge before contact and those that don't depend
strongly both on the compactness and on currently uknown higher PN corrections
to tidal effects. 

To end this subsection, let us mention that our results are {\it robust} under the choice of the EOB 
parameters $a_5$ and $a_6$ entering the BBH radial $A^0 (r)$ potential, Eq.~(\ref{eq:3.4}). The 
comparison between the currently most sophisticated version of the EOB formalism and the most 
accurate numerical relativity simulations has constrained the couple of parameters $(a_5 , a_6)$ to 
lie within a rather thin banana-like region in the $(a_5 , a_6)$ plane. We have checked that the 
results that we present in this paper are quite insensitive to the choice of $a_5$ and $a_6$ within this 
``good'' region. The default values that we use in the present paper are $a_5 = -6.37$, $a_6 = +50$, 
which lie in the ``good'' region. To illustrate the insensitivity of our
results to this choice, let us mention that  the value of twice
the orbital frequency at LSO, $2M\Omega_{\rm LSO}^{\rm EOB} (a_5 , a_6)$
(for an equal mass BNS system and for $c=0.17$),
changes from the  value $0.13605$, quoted in Table~\ref{tab:table3}, to the
new value $0.13603$ for $a_5=-4$ and $a_6=24$ which lie near the upper boundary
of the good region of paramaters discussed in Ref.~\cite{Damour:2009kr}. 

\subsection{Phasing and waveform from the non-adiabatic inspiral of 
tidally interacting compact binaries}
\label{sec:phasing}

Let us now consider the motion and radiation of tidally interacting binaries predicted by the full EOB 
formalism, i.e. beyond the adiabatic approximation. This is obtained by integrating the EOB equations 
of motion
$$
\frac{dr}{dt} = a(r) \, \frac{\partial \hat H_{\rm EOB}}{\partial \, p_{R_*}} \, ,
$$
$$
\frac{dp_{r_*}}{dt} = - a(r) \, \frac{\partial \hat H_{\rm EOB}}{\partial \, r} \, ,
$$
$$
\frac{d\varphi}{dt} = \frac{\partial \hat H_{\rm EOB}}{\partial \, p_\varphi} \, ,
$$
\be
\label{eq:5.11}
\frac{dp_\varphi}{dt} = \hat{\mathcal F}_{\varphi} \, , 
\ee
where $a(r) \equiv AD^{-1/2}$, $\hat H_{\rm EOB} (r,p_{r_*} , p \varphi) \equiv H_{\rm EOB} / \mu$, 
with $H_{\rm EOB}$ defined by Eq.~(\ref{eq:Heob}) above, and where the (scaled) radiation reaction 
$\hat{\mathcal F}_{\varphi} = {\mathcal F}_{\varphi} / \mu$ is defined in the
way introduced in~\cite{Damour:2009vw} improved (see Eq.~(3) there), i.e. by
summing over $\ell$ and $m$ the adiabatic multipolar partial 
fluxes corresponding to the newly resummed multipolar waves $h_{\l m}$ (including the tidal 
correction (\ref{eq:4.8}) in $h_{22}$). In addition, we recall that $r \equiv R/GM$, $t \equiv T/GM$, 
$p_{\varphi} \equiv P_{\varphi} / GM\mu$, and that the function $A(r)$ is here defined as the sum 
(\ref{eq:3.5}). Concerning the other metric coefficient $D^{-1} (r)$ (entering the auxiliary function 
$a \equiv (A/B)^{1/2} \equiv AD^{-1/2}$) we replace it by its standard resummation $(u \equiv 1/r)$
\be
\label{eq:5.12}
D^{-1} (r) = 1+6 \, \nu \, u^2 + 2 \, (26 - 3 \, \nu) \, \nu \, u^3 \, .
\ee

The solution of the ODE's (\ref{eq:5.11}) is then inserted in the newly resummed (and tidally 
completed) multipolar waves $h_{\l m}$ to compute the waveform emitted by the inspiralling 
compact binary. Here, we shall focus on the $\l = 2$, $m=2$ dominant asymptotic waveform 
$\lim_{R \to \infty} (R \, h_{22})$. 
%\footnote{?? Contrary to our previous work we do 
%not scale here the metric waveform in the ``Zerilli'' way, i.e. $Z_{22} \equiv \Psi_{22} \equiv 
%Rh_{22} / \sqrt{24}$.} 
Scaling it by $G\mu \equiv GM\nu$ and decomposing it in amplitude and phase,
\be
\label{eq:5.13}
\frac{R}{GM} \, \frac{h_{22}}{\nu} = A_{22} (t) \, e^{-{\rm i} \phi_{22}(t)} \, ,
\ee
we can then consider the dominant ``metric'' gravitational wave frequency $\omega_{22}(t) 
\equiv d \, \phi_{22} (t) / dt$. [Note that all these quantities are
  dimensionless. In particular $\omega_{22} \equiv GM\omega_{22}^{\rm phys}$.]

Up to now we have discussed  an extension of the EOB formalism which 
incorporates tidal effects in both the motion and the radiation of 
compact binaries. However, it has been advocated~\cite{Flanagan:2007ix,Read:2009yp,Hinderer:2009}
to incorporate tidal effects as a modification of one of the non-resummed 
``post-Newtonian''-based ways of describing the dynamics of inspiralling
binaries. 
In particular, the recent Ref.~\cite{Hinderer:2009} uses as baseline a
time-domain T4-type incorporation of tidal effects.
To be precise, let us recall that the phasing of the T4 approximant is defined
by the following ODEs
\begin{align}
\frac{d\phi_{22}^{\rm T4}}{dt} &= 2 \, x^{3/2}, \nonumber\\
\label{eq:T4bis}
\frac{dx}{dt} &= \frac{64}{5} \, \nu \, x^5 \,\left\{ a_{3.5}^{\rm Taylor} (x)
+  a^{\rm tidal}(x)\right\}
\end{align}
where $a_{3.5}^{\rm Taylor}$ is the PN expanded expression describing
pojnt-mass contributions, and where $a^{\rm tidal}$ 
is given in the equal mass case by~\cite{Flanagan:2007ix}
\be
\label{t4:lo}
a^{\rm tidal}(x)=26 \, \k_2^{\rm T} x^5.
\ee
Here we shall analyze the (metric) GW phase $\phi_{22}$ as a function of
the corresponding dimensionless frequency $\omega_{22}$ and study the
influence on it of tidal effects. More precisely, we give here two different 
comparisons between the EOB predictions and the T4
one. In these two comparisons, we keep T4 unchanged and defined by
Eq.~\eqref{eq:T4bis}, with a tidal contribution of the {\it leading order} (LO)
type~\eqref{t4:lo}. On the other hand, we compare this tidal-T4 model to two
different tidal-EOB models; both models use a tidally modified $A$ function, 
Eq.~\eqref{eq:3.5}. One model ($\rm EOB^{LO}$) uses the LO $A^{\rm tidal}$,
Eq.~\eqref{eq:3.6}, while the other one ($\rm EOB^{NLO}$) uses 
the {\it  Taylor} NLO $A^{\rm tidal}$, Eq.~\eqref{eq:linear}, with $\bar{\alpha}_1=7$. 
Here we consider a BNS equal-mass system  modelled using the 2H EOS with
compactness $c=0.13097$, mass $M=1.35M_\odot$ and radius $R=15.23$ km.

\begin{figure}[t]
\begin{center}
\includegraphics[width=75 mm, height=60mm]{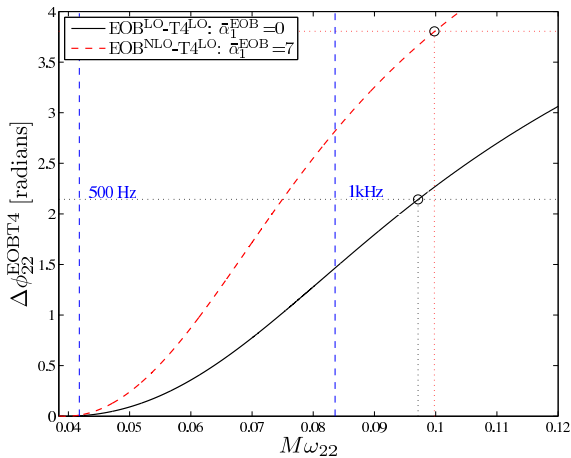}
\caption{\label{fig:fig5}Accumulated GW phase difference (versus GW frequency $\omega_{22}$) 
between tidal-EOB (quadrupolar) waveforms and a Taylor-T4-based PN 
waveform with (leading order) tidal corrections, Eq.~\eqref{eq:T4bis}.
Waveforms have been suitably aligned (subtracting a relative time and phase shift)
 at low frequencies. The circles on the plot indicate, for each curve, the  dephasing accumulated 
 up to the ``contact'' frequencies.}
  \end{center}
\end{figure}

The quantity which is plotted in Fig.~\ref{fig:fig5} is the difference 
$\Delta\phi_{22}^{\rm EOBT4}(\omega_{22})\equiv\phi_{22}^{\rm EOB^X}(\omega_{22})-\phi_{22}^{\rm T4}(\omega_{22})$.
where the label X on EOB takes two values, X=LO for the leading order model
and X=NLO for the next-to-leading order model.
To compute this quantity we took into account possible shifts in both 
$t \, (t^{\rm T4} = t^{\rm EOB} + \tau)$ and 
$\phi \, (\phi^{\rm T4} = \phi^{\rm EOB} + \alpha)$. We use here 
the ``two-frequency pinching'' technique of Ref.~\cite{Damour:2007vq} 
to fix suitable values of the shifts
$\tau$ and $\alpha$. We use here two pinching frequencies which are close to
$450$~Hz. In other words, the phase differences displayed in our figure show
the phase differences accumulated for frequencies between 450~Hz and the 
contact. Though, the figure does not display the phase differences below 450~Hz
we have checked that they remain much smaller than what they become for
frequencies higher than 450~Hz.

In Fig.~\ref{fig:fig5}, the solid line (black online) displays 
$\Delta\phi_{22}^{\rm EOBT4}(\omega_{22})=\phi_{22}^{\rm EOB^{LO}}(\omega_{22})-\phi_{22}^{\rm T4}(\omega_{22})$.
and the dashed line (red online) 
$\Delta\phi_{22}^{\rm EOBT4}(\omega_{22})=\phi_{22}^{\rm EOB^{NLO}}(\omega_{22})-\phi_{22}^{\rm T4}(\omega_{22})$.
%---------------------------------------
% FIG.5 EOB-T4 nonadiabatic comparisons.
%---------------------------------------
%--------------------
The two circles on the curves indicate the final moments of
``contact''. We added two vertical (dashed) lines corresponding 
to 500~Hz and 1~kHz.

The main messages that one can draw from this figure are: i) the relative
dephasing between EOB and T4 (using the same tidal model) grows by more 
than two radians up to contact; ii) the inclusion
of higher-order PN tidal contributions further increases
the relative dephasing by nearly two radians more.
Note that even if one stops the evolution around 1~kHz (which is within the
sensitivity of some possible configurations of Advanced LIGO) the previously
discussed accumulated dephasings are still larger than one radian. 
This indicates that the GW phasing of the ultimate part of the BNS inspiral
is very sensitive to tidal effects and also very sensitive to their precise
analytical modelling, including higher-order PN corrections.
This makes it urgent to do high-accuracy comparisons between accurate NR
simulations of BNS inspiral and EOB models, so as to accurately ``calibrate''
the EOB description of higher-order PN tidal contributions. 

%--------------------------
\section{Conclusions}
\label{conclusions}
%---------------------------

We discussed an extension of the EOB formalism which includes tidal effects.
The hope is that such a ``tidal-EOB'' formalism will be able to go beyond
the present PN-based proposals whose validity is limited to the early
(lower-frequency) portion of the GW inspiral signal emitted by BNS systems.
This formalism allows naturally for the presence of higher-order PN
corrections to the leading (Newtonian) effects. We compared tidal-EOB 
predictions to recently computed numerical relativity data of
quasi-equilibrium circular BNS sequences~\cite{Uryu:2009ye}. 
We showed how to subtract tidal effects from NR data. 
Even after this subtraction, there remains a systematic
difference between the ``point-mass'' NR binding energy and its EOB (and PN)
analytical correspondant. We argue that this difference is due to unaccounted 
3PN-level effects linked to the imperfect satisfaction of the 
helical Killing vector condition (which should be satisfied for physically  
waveless solutions). We advocate that new nonconformally flat simulations 
be performed for
sequences of helical-Killing-vector cut-off radii so as to allow
extrapolation to infinite radius.
We also suggested to study BHNS circular binaries for mass ratios
$M_{\rm BH}/M_{\rm NS}$ of order unity.

In absence of such physically waveless NR data, we propose to subtract from
the current data a term $\delta\,x^4$ representing a 3PN correction in the
binding energy. We could then do a least-square analysis to try to minimize
the (squared) ``distance'' $\chi^2$ between NR data and tidal-EOB predictions.
Our analysis allowed for 1PN-corrections to tidal effects parametrized
by $\bar{\alpha}_1$. We found that $\chi^2$ remains close to its global
minimum in a flat valley that extends over a significant region of the
$\left(\bar{\alpha}_1,\delta\right)$  plane. 
This means that, given the present error level in numerical data, we
cannot meaningfully and simultaneously select preferred values for 
$\bar{\alpha}_1$ and $\delta$.
Though this analysis is not fully conclusive, it does suggest the need
of including higher-order PN correction to tidal effects that {\it significantly}
increase their dynamical effect.
[In other words, the ``effective'' value, say 
$\k_2^{\rm eff}(u) =  \k_2^{\rm T}\left(1+\bar{\alpha}_1 u +\bar{\alpha}_2 u^2 + \dots\right)$, 
which is relevant for the late inspiral is significantly larger, by a
factor~$\sim 2$, than $\k_2^{\rm T}$].
These higher-order PN corrections might
come not from the 1PN level, but from higher PN levels (see in particular
the end of Sec.~\ref{sec:nr}, where a 2PN completion of a recently computed 
1PN correction of order unity was shown to be  fully compatible with  
current NR data).

This emphasizes the need both of higher order analytical calculations
of tidal effects and of high-accuracy numerical relativity simulations
of inspiralling BNS systems.[We note in this respect that it would be
useful to refine the piece-wise polytropic approximation to realistic
tabulated EOS (used in this paper) by incorporating the relativistic
Love numbers, notably  $k_2$, within the set of observables
that are fitted].
We argued that such a suitably tidally completed EOB formalism will
be able to describe the dynamics (and GW emission) of inspiralling
BNS systems essentially up to the contact of the two neutron stars.
We emphasized that, though below the dimensionless (quadrupolar) 
GW frequency $GM\omega_{22}\sim 0.04$ (which corresponds to a 
freqency of 480~Hz for $1.35M_\odot +1.35M_\odot$ system) the present
analytical knowledge is possibly sufficient for accurately describing the
system, the GW phasing becomes uncertain by a large amount ($\sim 4 $~radians)
during the late part of the inspiral,
because of our current lack of secure knowledge of higher order PN
corrections to tidal effects.
This makes it urgent to do high-accuracy comparisons between accurate NR
simulation of BNS inspiral and EOB models.
When the EOB description of higher-PN tidal effects is ``calibrated'' with sufficient
accuracy by using such EOB/NR comparisons, we think it will be possible
to use the EOB formalism to extract from Advanced-LIGO data some accurate 
knowledge of the nuclear EOS (via the measurement of the crucial parameter
$\k_2^{\rm T}$). 

\acknowledgments
We are grateful to Lo\"ic Villain for collaboration, at an early stage, on
the NR-EOB comparison. We thank Koji Ury${\rm\bar{u}}$ for making available to us the
numerical data behind the published tables of Ref.~\cite{Uryu:2009ye}.
We are also grateful to Luca Baiotti, Bruno Giacomazzo and 
Luciano Rezzolla for sharing with us, before publication, 
their data on inspiralling and coalescing binary neutron
stars, which prompted our interest in relativistic tidal 
properties of neutron stars.
\appendix


\begin{thebibliography}{200}

\bibitem{Damour:2009kr}
  T.~Damour and A.~Nagar,
  %``An improved analytical description of inspiralling and coalescing
  %black-hole binaries,''
  Phys.\ Rev.\  D {\bf 79}, 081503 (2009)
  [arXiv:0902.0136 [gr-qc]].
  %%CITATION = PHRVA,D79,081503;%%


\bibitem{Buonanno:2009qa}
  A.~Buonanno, Y.~Pan, H.~P.~Pfeiffer, M.~A.~Scheel, L.~T.~Buchman and L.~E.~Kidder,
  %``Effective-one-body waveforms calibrated to numerical relativity
  %simulations: coalescence of non-spinning, equal-mass black holes,''
  Phys.\ Rev.\  D {\bf 79}, 124028 (2009)
  [arXiv:0902.0790 [gr-qc]].
  %%CITATION = PHRVA,D79,124028;%%


\bibitem{Yunes:2009ef}
  N.~Yunes, A.~Buonanno, S.~A.~Hughes, M.~C.~Miller and Y.~Pan,
  %``Modeling Extreme Mass Ratio Inspirals within the Effective-One-Body
  %Approach,''
  arXiv:0909.4263 [gr-qc].
  %%CITATION = ARXIV:0909.4263;%%

%\cite{Baiotti:2009gk}
\bibitem{Baiotti:2009gk}
  L.~Baiotti, B.~Giacomazzo and L.~Rezzolla,
  %``Accurate evolutions of inspiralling neutron-star binaries: assessment of
  %the truncation error,''
  Class.\ Quant.\ Grav.\  {\bf 26}, 114005 (2009)
  [arXiv:0901.4955 [gr-qc]].
  %%CITATION = CQGRD,26,114005;%%

%\cite{Giacomazzo:2009mp}
\bibitem{Giacomazzo:2009mp}
  B.~Giacomazzo, L.~Rezzolla and L.~Baiotti,
  %``Can magnetic fields be detected during the inspiral of binary neutron
  %stars?,''
  Mon.\ Not.\ Roy.\ Astron.\ Soc.\  {\bf 399}, L164 (2009)
  [arXiv:0901.2722 [gr-qc]].
  %%CITATION = MNRAA,399,L164;%

%\cite{Baiotti:2008ra}
\bibitem{Baiotti:2008ra}
  L.~Baiotti, B.~Giacomazzo and L.~Rezzolla,
  %``Accurate evolutions of inspiralling neutron-star binaries: prompt and
  %delayed collapse to black hole,''
  Phys.\ Rev.\  D {\bf 78}, 084033 (2008)
  [arXiv:0804.0594 [gr-qc]].
  %%CITATION = PHRVA,D78,084033;%%


\bibitem{Bildsten:1992my}
  L.~Bildsten and C.~Cutler,
  %``Tidal interactions of inspiraling compact binaries,''
  Astrophys.\ J.\  {\bf 400}, 175 (1992).
  %%CITATION = ASJOA,400,175;%%


\bibitem{Kochanek:1992wk}
  C.~S.~Kochanek,
  %``Coalescing binary neutron stars,''
  Astrophys.\ J.\  {\bf 398}, 234 (1992).
  %%CITATION = ASJOA,398,234;%%


\bibitem{Vallisneri:1999nq}
  M.~Vallisneri,
  %``Prospects for gravitational-wave observations of neutron-star tidal
  %disruption in neutron-star/black-hole binaries,''
  Phys.\ Rev.\ Lett.\  {\bf 84}, 3519 (2000)
  [arXiv:gr-qc/9912026].
  %%CITATION = PRLTA,84,3519;%%


\bibitem{Flanagan:2007ix}
  E.~E.~Flanagan and T.~Hinderer,
  %``Constraining neutron star tidal Love numbers with gravitational wave
  %detectors,''
  Phys.\ Rev.\  D {\bf 77}, 021502 (2008)
  [arXiv:0709.1915 [astro-ph]].
  %%CITATION = PHRVA,D77,021502;%%


\bibitem{Hinderer:2007mb}
  T.~Hinderer,
  %``Tidal Love numbers of neutron stars,''
  Astrophys.\ J.\  {\bf 677}, 1216 (2008)
  [arXiv:0711.2420 [astro-ph]].
  %%CITATION = ASJOA,677,1216;%%

\bibitem{Hinderer:2009}
  T.~Hinderer, B.~D.~Lackey, R.~N.~Lang and J.~S.~Read,
  %``Tidal deformability of neutron stars with realistic equations of state and
  %their gravitational wave signatures in binary inspiral,''
  arXiv:0911.3535 [astro-ph.HE].
  %%CITATION = ARXIV:0911.3535;%%

\bibitem{Damour:2009vw}
  T.~Damour and A.~Nagar,
  %``Relativistic tidal properties of neutron stars,''
  Phys.\ Rev.\  D {\bf 80}, 084035 (2009)
  [arXiv:0906.0096 [gr-qc]].
  %%CITATION = PHRVA,D80,084035;%%


\bibitem{Binnington:2009bb}
  T.~Binnington and E.~Poisson,
  %``Relativistic theory of tidal Love numbers,''
  Phys.\ Rev.\  D {\bf 80}, 084018 (2009)
  [arXiv:0906.1366 [gr-qc]].
  %%CITATION = PHRVA,D80,084018;%%

\bibitem{Uryu:2009ye}
  K.~Ury${\rm\bar{u}}$, F.~Limousin, J.~L.~Friedman, E.~Gourgoulhon and M.~Shibata,
  %``Non-conformally flat initial d$\Omega_{\rm LSO}^{\rm EOB} (a_5 , a_6)$ata for binary compact objects,''
  arXiv:0908.0579 [gr-qc].
  %%CITATION = ARXIV:0908.0579;%%

%\cite{Uryu:2005vv}
\bibitem{Uryu:2005vv}
  K.~Ury${\rm\bar{u}}$, F.~Limousin, J.~L.~Friedman, E.~Gourgoulhon and M.~Shibata,
  %``Binary neutron stars in a waveless approximation,''
  Phys.\ Rev.\ Lett.\  {\bf 97}, 171101 (2006)
  [arXiv:gr-qc/0511136].
  %%CITATION = PRLTA,97,171101;%%

\bibitem{Damour_cras80}
T.~Damour, 
%``Masses ponctuelles en Relativit\'e g\'en\'erale'',
C. R. Acad. Sc. Paris, S\'erie A, {\bf 291}, 227 (1980).

\bibitem{Damour1983}
T.~Damour, in {\it Gravitational Radiation}, edited by N.~Deruelle and
T.~Piran (North-Holland, Amsterdam, 1983), p.59;

\bibitem{Damour:2001bu}
  T.~Damour, P.~Jaranowski and G.~Schaefer,
  %``Dimensional regularization of the gravitational interaction of point
  %masses,''
  Phys.\ Lett.\  B {\bf 513}, 147 (2001)
  [arXiv:gr-qc/0105038].
  %%CITATION = PHLTA,B513,147;%%

\bibitem{Blanchet:2003gy}
  L.~Blanchet, T.~Damour and G.~Esposito-Farese,
  %``Dimensional regularization of the third post-Newtonian dynamics of point
  %particles in harmonic coordinates,''
  Phys.\ Rev.\  D {\bf 69}, 124007 (2004)
  [arXiv:gr-qc/0311052].
  %%CITATION = PHRVA,D69,124007;%%

\bibitem{Blanchet:2004ek}
  L.~Blanchet, T.~Damour, G.~Esposito-Farese and B.~R.~Iyer,
  %``Gravitational radiation from inspiralling compact binaries completed at
  %the third post-Newtonian order,''
  Phys.\ Rev.\ Lett.\  {\bf 93}, 091101 (2004)
  [arXiv:gr-qc/0406012].
  %%CITATION = PRLTA,93,091101;%%

\bibitem{Damour:1995kt}
  T.~Damour and G.~Esposito-Farese,
  %``Testing gravity to second postNewtonian order: A Field theory approach,''
  Phys.\ Rev.\  D {\bf 53}, 5541 (1996)
  [arXiv:gr-qc/9506063].
  %%CITATION = PHRVA,D53,5541;%%

%\cite{Goldberger:2004jt}
\bibitem{Goldberger:2004jt}
  W.~D.~Goldberger and I.~Z.~Rothstein,
  %``An effective field theory of gravity for extended objects,''
  Phys.\ Rev.\  D {\bf 73}, 104029 (2006)
  [arXiv:hep-th/0409156].
  %%CITATION = PHRVA,D73,104029;%%

%\cite{Damour:1998jk}
\bibitem{Damour:1998jk}
  T.~Damour and G.~Esposito-Farese,
  %``Gravitational-wave versus binary-pulsar tests of strong-field gravity,''
  Phys.\ Rev.\  D {\bf 58}, 042001 (1998)
  [arXiv:gr-qc/9803031].
  %%CITATION = PHRVA,D58,042001;%%


\bibitem{Damour:1990pi}
  T.~Damour, M.~Soffel and C.~m.~Xu,
  %``General relativistic celestial mechanics. 1. Method and definition of
  %reference systems,''
  Phys.\ Rev.\  D {\bf 43}, 3272 (1991).
  %%CITATION = PHRVA,D43,3272;%%


\bibitem{Damour:2009sm}
  T.~Damour,
  %``Gravitational Self Force in a Schwarzschild Background and the Effective
  %One Body Formalism,''
  arXiv:0910.5533 [gr-qc].
  %%CITATION = ARXIV:0910.5533;%%


\bibitem{BD86}
 L.~Blanchet and T.~Damour,
Phil. \ Trans.\ R. Soc. Lond. A, {\bf 320}, 379 (1986).

\bibitem{Damour:1991yw}
  T.~Damour, M.~Soffel and C.~m.~Xu,
  %``General relativistic celestial mechanics. 2. Translational equations of
  %motion,''
  Phys.\ Rev.\  D {\bf 45}, 1017 (1992).
  %%CITATION = PHRVA,D45,1017;%%


\bibitem{DEF09}
T.~Damour and G.~Esposito-Far\`ese, in preparation


\bibitem{Damour:1992qi}
  T.~Damour, M.~Soffel and C.~m.~Xu,
  %``General Relativistic Celestial Mechanics. 3. Rotational Equations Of
  %Motion,''
  Phys.\ Rev.\  D {\bf 47}, 3124 (1993).
  %%CITATION = PHRVA,D47,3124;%%

\bibitem{Damour:1993zn}
  T.~Damour, M.~Soffel and C.~m.~Xu,
  %``General relativistic celestial mechanics. 4: Theory of satellite motion,''
  Phys.\ Rev.\  D {\bf 49}, 618 (1994).
  %%CITATION = PHRVA,D49,618;%%


\bibitem{Buonanno:1998gg}
  A.~Buonanno and T.~Damour,
  %``Effective one-body approach to general relativistic two-body dynamics,''
  Phys.\ Rev.\  D {\bf 59}, 084006 (1999)
  [arXiv:gr-qc/9811091].
  %%CITATION = PHRVA,D59,084006;%%


\bibitem{Buonanno:2000ef}
  A.~Buonanno and T.~Damour,
  %``Transition from inspiral to plunge in binary black hole coalescences,''
  Phys.\ Rev.\  D {\bf 62}, 064015 (2000)
  [arXiv:gr-qc/0001013].
  %%CITATION = PHRVA,D62,064015;%%

\bibitem{Damour:2001tu}
  T.~Damour,
  %``Coalescence of two spinning black holes: An effective one-body  approach,''
  Phys.\ Rev.\  D {\bf 64}, 124013 (2001)
  [arXiv:gr-qc/0103018].
  %%CITATION = PHRVA,D64,124013;%%

\bibitem{Damour:2000we}
  T.~Damour, P.~Jaranowski and G.~Schaefer,
  %``On the determination of the last stable orbit for circular general
  %relativistic binaries at the third post-Newtonian approximation,''
  Phys.\ Rev.\  D {\bf 62}, 084011 (2000)
  [arXiv:gr-qc/0005034].
  %%CITATION = PHRVA,D62,084011;%%


\bibitem{Damour:2009ic}
  T.~Damour and A.~Nagar,
  %``The Effective One Body description of the Two-Body problem,''
  arXiv:0906.1769 [gr-qc].
  %%CITATION = ARXIV:0906.


\bibitem{Mora:2003wt}
  T.~Mora and C.~M.~Will,
  %``A Post-Newtonian diagnostic of quasi-equilibrium binary configurations of
  %compact objects,''
  Phys.\ Rev.\  D {\bf 69}, 104021 (2004)
  [Erratum-ibid.\  D {\bf 71}, 129901 (2005)]
  [arXiv:gr-qc/0312082].
  %%CITATION = PHRVA,D69,104021;%%


\bibitem{Damour:1999cr}
  T.~Damour, P.~Jaranowski and G.~Schaefer,
  %``Dynamical invariants for general relativistic two-body systems at the third
  %post-Newtonian approximation,''
  Phys.\ Rev.\  D {\bf 62}, 044024 (2000)
  [arXiv:gr-qc/9912092].
  %%CITATION = PHRVA,D62,044024;%%


\bibitem{Read:2008iy}
  J.~S.~Read, B.~D.~Lackey, B.~J.~Owen and J.~L.~Friedman,
  %``Constraints on a phenomenologically parameterized neutron-star equation of
  %state,''
  Phys.\ Rev.\  D {\bf 79}, 124032 (2009)
  [arXiv:0812.2163 [astro-ph]].
  %%CITATION = PHRVA,D79,124032;%%


\bibitem{Read:2009yp}
  J.~S.~Read, C.~Markakis, M.~Shibata, K.~Ury${\rm\bar{u}}$, J.~D.~E.~Creighton and J.~L.~Friedman,
  %``Measuring the neutron star equation of state with gravitational wave
  %observations,''
  Phys.\ Rev.\  D {\bf 79}, 124033 (2009)
  [arXiv:0901.3258 [gr-qc]].
  %%CITATION = PHRVA,D79,124033;%%


\bibitem{Jaranowski:1997ky}
  P.~Jaranowski and G.~Schaefer,
  %``3rd post-Newtonian higher order Hamilton dynamics for two-body point-mass
  %systems,''
  Phys.\ Rev.\  D {\bf 57}, 7274 (1998)
  [Erratum-ibid.\  D {\bf 63}, 029902 (2001)]
  [arXiv:gr-qc/9712075].
  %%CITATION = PHRVA,D57,7274;%%


\bibitem{Damour:2008gu}
  T.~Damour, B.~R.~Iyer and A.~Nagar,
  %``Improved resummation of post-Newtonian multipolar waveforms from
  %circularized compact binaries,''
  Phys.\ Rev.\  D {\bf 79}, 064004 (2009)
  [arXiv:0811.2069 [gr-qc]].
  %%CITATION = PHRVA,D79,064004;%%


\bibitem{Damour:2006tr}
  T.~Damour and A.~Gopakumar,
  %``Gravitational recoil during binary black hole coalescence using the
  %effective one body approach,''
  Phys.\ Rev.\  D {\bf 73}, 124006 (2006)
  [arXiv:gr-qc/0602117].
  %%CITATION = PHRVA,D73,124006;%%

\bibitem{Damour:2007vq}
  T.~Damour, A.~Nagar, E.~N.~Dorband, D.~Pollney and L.~Rezzolla,
  %``Faithful Effective-One-Body waveforms of equal-mass coalescing black-hole
  %binaries,''
  Phys.\ Rev.\  D {\bf 77}, 084017 (2008)
  [arXiv:0712.3003 [gr-qc]].
  %%CITATION = PHRVA,D77,084017;%%

\bibitem{Damour:2007yf}
  T.~Damour and A.~Nagar,
  %``Comparing Effective-One-Body gravitational waveforms to accurate numerical
  %data,''
  Phys.\ Rev.\  D {\bf 77}, 024043 (2008)
  [arXiv:0711.2628 [gr-qc]].
  %%CITATION = PHRVA,D77,024043;%%

\end{thebibliography}
\end{document}